\documentclass[a4paper,11pt]{article}
\usepackage[utf8]{inputenc}
\usepackage[english]{babel}
\usepackage{braket}
\usepackage{xspace}
\usepackage{bbm}
\usepackage{mathdots}
\usepackage{stackrel}
\usepackage[dvipsnames]{xcolor}
\usepackage{appendix}
\usepackage{hyperref}
\hypersetup{
 colorlinks=true,
 linkcolor=blue,
 anchorcolor = blue,
 citecolor = blue,
 filecolor = blue,
 urlcolor = blue
}
\usepackage{mathtools}
\usepackage{bbm}
\usepackage{comment}
\usepackage{subcaption}
\usepackage[T1]{fontenc}               % codifica dei font
\usepackage[left=3cm,
            right=2.5cm,
            top=2.5cm,
            bottom=3cm
            ]{geometry}                % spaziatura bordi
\usepackage{graphicx}
\usepackage{mathrsfs}
\usepackage{frcursive}
\usepackage{appendix}
\usepackage{fancyhdr}
\usepackage{amsmath,                   % simboli matematici
            amssymb,                   % altri simboli matematici
            amsthm}                    % stili teorema
\usepackage[capitalise]{cleveref}
\usepackage{setspace}
\usepackage{cite}
\usepackage{cancel}
\usepackage{bm}
\usepackage{mathrsfs}
\usepackage{tikz}
\usetikzlibrary{arrows}
\makeatletter
\newcommand\@erelb@r[1]{%
	\mathrel{\tikz[baseline=-.5ex]\draw[#1] (0,0)--(0.3,0);}
}
\newcommand{\erelbar}[1]{\@erelbar#1}
\def\@erelbar#1#2{%
	\ifcase\numexpr#1*4+#2\relax
	\@erelb@r{-}\or     % 00
	\@erelb@r{->}\or    % 01
	\@erelb@r{-|}\or    % 02
	\@erelb@r{->|}\or   % 03
	\@erelb@r{<-}\or    % 10
	\@erelb@r{<->}\or   % 11
	\@erelb@r{<-|}\or   % 12
	\@erelb@r{<->}\or   % 13
	\@erelb@r{|-}\or    % 20
	\@erelb@r{|->}\or   % 21
	\@erelb@r{|-|}\or   % 22
	\@erelb@r{|<->|}\or % 23
	\@erelb@r{|<-}\or   % 30
	\@erelb@r{|<->}\or  % 31
	\@erelb@r{|<-|}\or  % 32
	\@erelb@r{|<->|}    % 33
	\else
	\@wrong
	\fi
}
\makeatother
\usepackage[
margin=30pt,
labelfont=bf,
font=small,
justification=justified,   % Fully justified caption
singlelinecheck=false       % Important: do not center short captions
]{caption}

\usepackage{setspace} 
\usepackage{scalerel}
\usetikzlibrary{svg.path}
\definecolor{orcidlogocol}{HTML}{A6CE39}
\tikzset{
  orcidlogo/.pic={
    \fill[orcidlogocol] svg{M256,128c0,70.7-57.3,128-128,128C57.3,256,0,198.7,0,128C0,57.3,57.3,0,128,0C198.7,0,256,57.3,256,128z};
    \fill[white] svg{M86.3,186.2H70.9V79.1h15.4v48.4V186.2z}
                 svg{M108.9,79.1h41.6c39.6,0,57,28.3,57,53.6c0,27.5-21.5,53.6-56.8,53.6h-41.8V79.1z M124.3,172.4h24.5c34.9,0,42.9-26.5,42.9-39.7c0-21.5-13.7-39.7-43.7-39.7h-23.7V172.4z}
                 svg{M88.7,56.8c0,5.5-4.5,10.1-10.1,10.1c-5.6,0-10.1-4.6-10.1-10.1c0-5.6,4.5-10.1,10.1-10.1C84.2,46.7,88.7,51.3,88.7,56.8z};
  }
}

\newcommand\orcidicon[1]{\href{https://orcid.org/#1}{\mbox{\scalerel*{
\begin{tikzpicture}[yscale=-1,transform shape]
\pic{orcidlogo};
\end{tikzpicture}
}{|}}}}

\title{\bf A closed system setting for quantum thermalisation in free fermions}

\author{Purvaash Panduranghan-Udhayashankar$^1$, Filiberto Ares$^1$, Pasquale Calabrese$^{1}$}

\date{}

\begin{document} 

\maketitle

{\small
\vspace{-5mm}  \ \\
{$^{1}$}  SISSA and INFN, via Bonomea 265, 34136 Trieste, Italy\\[-0.1cm]
\medskip
}

\begin{abstract}

We study thermalisation and the possible occurrence of the Mpemba effect in a closed quantum setting that mimics the interaction of a system with thermal reservoirs coupled only at its boundaries. Specifically, we consider a tripartite geometry in which a finite chain, initially prepared at a finite temperature, is suddenly connected on both sides to two semi-infinite chains of the same nature held at a different temperature. These outer chains act as thermal baths, while the full system evolves unitarily under a homogeneous Hamiltonian.
This setup provides a simple quantum realisation of a temperature quench and closely resembles the original scenario in which the classical Mpemba effect was first observed. We focus on two paradigmatic free-fermion models, namely the XX chain and the transverse-field Ising chain, which respectively preserve and break the global $U(1)$ particle-number symmetry.
As a probe of relaxation, we consider the Frobenius distance between the time-evolved reduced density matrix of the central subsystem and its stationary state, which, as expected, is the thermal state at the bath temperature. Exploiting the free-fermionic structure of both models, the dynamics remains Gaussian and the Frobenius distance can be expressed exactly in terms of two-point correlation functions. Combining this representation with generalised hydrodynamics, we derive analytical predictions for the Frobenius distance in the hydrodynamic limit, providing a complete characterisation of the thermalisation process.
Using these results, we investigate the possible occurrence of the Mpemba effect. We find that, despite the genuine non-equilibrium dynamics displayed by the system, no Mpemba effect arises in this setting. Our analysis therefore identifies a broad class of boundary-driven thermalisation protocols in which relaxation is fully characterised analytically and exhibits no anomalous acceleration of equilibration.

\end{abstract}
\newpage 

\tableofcontents

\section{Introduction}

A fundamental problem in statistical physics is understanding what happens when a system is brought into contact with a thermal bath at a different  temperature. The laws of thermodynamics tell us that 
the system will eventually relax to equilibrium and attain the temperature of the bath. However, understanding the thermalisation process itself (that is, how the system approaches equilibrium and on what timescales) is a considerably more challenging problem. 
For example, one might intuitively expect that the larger the temperature difference between the 
system and the bath, the longer it takes for the system to equilibrate. Yet this is not necessarily the case: a hotter system can freeze faster than a colder one~\cite{mpemba-69}. This is a manifestation of the Mpemba effect, a genuinely out-of-equilibrium phenomenon, in which a system prepared further from equilibrium may relax faster than a system closer to equilibrium, depending on the  underlying dynamics~\cite{tbl-26}. This effect has been observed in many different systems~\cite{akkl-16, hlhl-18, cdkb-10, glcg-11, lasanta-17, keller-18, ths-21, santos-24, lvvcwh-26} and is particularly well understood in classical Markovian systems~\cite{lr-17, krhv-19, bkc-21, tyr-23, vvh-25, wv-22}, where theoretical predictions have also been corroborated experimentally~\cite{kb-20}, and where the inverse Mpemba effect has also been identified~\cite{lr-17, kcb-22}.

The Mpemba effect has also recently been observed in quantum systems~\cite{acm-25}. In this context, two main classes of 
physical settings have been investigated. One involves 
open quantum systems whose dynamics are governed by a 
Lindblad equation describing the coupling to an 
environment~\cite{nv-19, cll-21, cth-23, moroder-24, ne-24, spc-25, ne-25, longhi-25, wpm-25, bt-26}. Here, the Mpemba effect can be explained 
in terms of the overlap between the initial state and 
the slowest mode of the Lindbladian. The other scenario 
concerns closed quantum systems undergoing unitary 
evolution, i.e. quantum quenches~\cite{amc-23, calabrese-26}. In this second case, 
the full system does not relax to equilibrium. However, 
if one considers a subsystem, then in the thermodynamic 
limit the rest of the system effectively acts as a bath 
and the subsystem equilibrates to a statistical 
ensemble. An explanation of the quantum Mpemba effect 
in closed systems has been given in terms of 
quasiparticle transport in integrable 
systems~\cite{amc-23,calabrese-26, rylands-24, makc-24, bertini-24, chalas-24, rvc-24, klobas-24, yac-24, yca-25, arc-25, yet-25, pa-26, tc-26,se-26,ggb-25}, while in chaotic quantum systems, several 
model-specific mechanisms have been proposed~\cite{liu-24, tcdl-25, foligno-25, ylz-25, yu-25-rev, yjzxf-25, bhore-25, ya-26, mcroberts-26, yh-26, mpf-26, asst-25}. See also Refs.~\cite{cma-24, avm-25, dgtm-25, rsca-26, bds-25, benini-25, liu-mb-26, krb-25} for other different systems and settings. In general, all these works consider quantum quenches from spatially homogeneous initial states, which are not necessarily pure.
Both the 
open and closed cases have also been realized 
experimentally in quantum simulators~\cite{Joshi-24, shapira-24, zhang-25, xfc-25, xia-26}.

In the present work, we investigate a different route to the thermalisation of a quantum system that remains entirely unitary. Our approach is based on a tripartite quantum quench designed to mimic the original setting in which the classical Mpemba effect was observed. Specifically, we consider a finite quantum spin chain initially prepared at a finite temperature and suddenly connected to two semi-infinite chains of the same type (i.e. governed by the same Hamiltonian) but prepared at a different temperature, which effectively act as thermal baths. The initial state is therefore spatially inhomogeneous. 
The full system then evolves unitarily under the same Hamiltonian. We study the final equilibrium state of the subsystem in contact with the baths and its relaxation dynamics toward that state. In particular, we investigate whether the Mpemba effect is present in this setting. 
Several quantities can be used to monitor the relaxation of the subsystem and probe the possible occurrence of the Mpemba effect. These include distances, relative entropies, and fidelities between the time-evolved state and its stationary counterpart, as well as quantum-resource measures such as the asymmetry, which characterises the dynamical restoration of symmetries during the evolution (see Ref.~\cite{acm-25} for a review of various distance measures and Ref.~\cite{smbtmg-26} for a unified resource-theoretic perspective).
Here, we consider the normalised Frobenius distance between the time-evolved state of the subsystem and the stationary state~\cite{fagotti-rdm-frob}, which has also been used as a probe in the experiment of Ref.~\cite{Joshi-24}.

We study two paradigmatic quantum spin-$1/2$ chains: 
the XX and transverse-field Ising models. These systems 
have the advantage that, through a Jordan-Wigner transformation, 
they can be mapped exactly onto free-fermionic chains 
and are therefore exactly solvable. As a consequence, 
the state of the system remains Gaussian at all times 
and is uniquely determined by its two-point correlation 
functions. This allows us to compute the Frobenius 
distance efficiently, even for large system sizes. 
Moreover, we can apply generalised hydrodynamics (GHD)~\cite{cdy-16, bcdf-16}, 
a systematic framework for studying quenches in 
integrable systems with inhomogeneous initial 
conditions such as the one considered here~\cite{bbdv-22, essler-22, doyon-20, dgmsv-25}. Within this 
framework, the state of the system at large space–time 
scales can be described locally in terms of a 
quasiparticle momentum occupation function evolving 
according to a continuity equation. GHD provides 
predictions for the evolution of conserved charges and 
currents~\cite{pdncbf-17, bulchandani-17, ddky-17, cdlv-18, dyc-18, bac-19, cddky-19}, which have been validated experimentally~\cite{sbdd-19, malvania-21,  dtdb-26}, 
and, in some cases, also for entanglement entropies~\cite{bfpc-18, abf-19, abfpr-21}. 
Here, we show that GHD can be used to obtain an exact 
expression for the Frobenius distance between the time-
evolved state of the subsystem and its stationary state 
in the hydrodynamic limit.

The paper is organised as follows. In 
Sec.~\ref{subsec:quench-setup-xx}, we introduce the 
tripartite quench, the Frobenius distance, and the 
methods for computing it. To our knowledge, the 
Frobenius distance has not yet been used as a probe of the 
Mpemba effect in free fermionic systems. For this reason, in 
Sec.~\ref{subsec:frobdist-qpp} we first consider a 
homogeneous quench from the ground state of the XY 
chain to the XX chain, comparing the results with those 
previously obtained from other observables. We also 
derive an exact expression for the Frobenius distance 
within the quasiparticle picture, which is useful 
for obtaining the predictions in the tripartite quench 
employing GHD. This is done in Sec.~\ref{sec:xx}, where 
we study the XX chain. In Sec.~\ref{sec:TFIC}, we repeat the same 
analysis for the Ising chain. Finally, we present our 
conclusions in Sec.~\ref{sec:conclusions}. We include Appendices~\ref{app:obc} and~\ref{app:ev_TIC} 
containing some technical details of the Ising case.

\section{Tripartite quench setup}
\label{subsec:quench-setup-xx}

\begin{figure}
	\centering
    \includegraphics[width=0.7\textwidth]{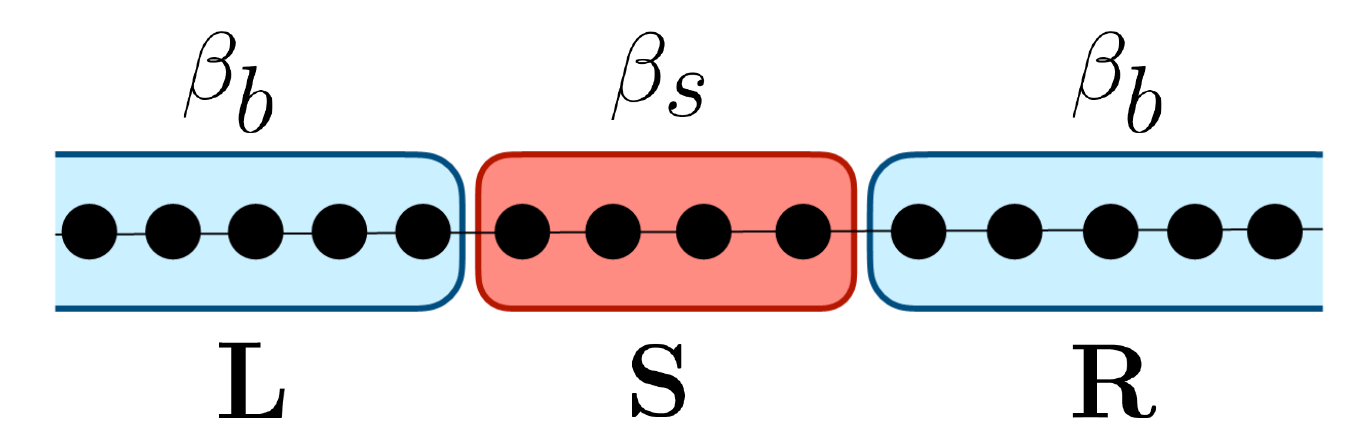}
	\caption{Tripartite quench setup studied in this paper. We initially prepare an open spin chain $S$ of size $\ell$ at temperature $1/\beta_s$. At $t=0$, it is connected at its ends to two semi-infinite chains $L$ and $R$ at  temperature $1/\beta_b$. At time $t>0$, the combined system evolves unitarily with a translationally invariant Hamiltonian.}
	\label{fig:tripartition}
\end{figure}

In this section, we introduce the tripartite quench we study and the basic tools we employ.
Our quench setup comprises the tripartite system shown in Fig.~\ref{fig:tripartition}. At $t=0$, the subsystem $S$ is an open spin-$1/2$ chain with $\ell$ sites at temperature $1/\beta_s$, and $L$ and $R$ are semi-infinite spin-$1/2$ chains at temperature $1/\beta_b$, which will act as reservoirs.
The initial state of the total system is then given by the tensor product of the state of each subsystem,
\begin{equation}
		\rho(t=0) = \rho_L\otimes \rho_S\otimes\rho_R,
\end{equation}
which are described by the Gibbs ensembles
\begin{align}
		\rho_L = \dfrac{e^{-\beta_b H_L}}{Z_L}, \quad
		\rho_S = \dfrac{e^{-\beta_s H_S}}{Z_s},\quad 
		\rho_R &= \dfrac{e^{-\beta_b H_R}}{Z_R}.
	\label{eq:xx_inital_state}
\end{align}
where $Z_\mu={\rm Tr}(e^{-\beta H_\mu})$ are the partition functions of each subsystem.
At $t>0$, we connect the endpoints of $S$ with the reservoirs $L$ and $R$ as shown in the figure and let the full system evolve unitarily, 
\begin{equation}\label{eq:evolved_state}
\rho(t)=e^{-itH}\rho(0)e^{itH},
\end{equation}
with a Hamiltonian $H=H_L+H_S+H_R+H_{LS}+H_{SR}$, where $H_{LS}$ and $H_{RS}$ contain the couplings between $S$ and the reservoirs $L$ and $R$. In our setup, both 
$S$ and the reservoirs 
$L$ and 
$R$ are described by the same spatially homogeneous Hamiltonian, either the XX spin chain or the transverse-field quantum Ising chain, and the post-quench Hamiltonian 
$H$ is likewise translationally invariant. 

Bipartite quenches, in which two semi-infinite chains at different temperatures are connected, have been extensively studied, as they provide the minimal setting for studying transport, including the XX and Ising spin chains~\cite{ogata-02, pk-05, pk-07, lvbd-13, dlmv-14, ez-13, kormos-17, perfetto-18, kz-17, bfpc-18, abfpr-21}. In that case, the system tends to a non-equilibrium steady state characterised by a generalised Gibbs ensemble with non-zero currents. To our knowledge, tripartite quenches such as the one shown in Fig.~\ref{fig:tripartition} have received considerably less attention, mostly focusing on the stationary-state properties when the temperatures of $L$ and $R$ are different~\cite{ap-03, ab-06}.

The goal of the paper is to study how the subsystem $S$ relaxes towards the stationary state in the long time limit. The state of $S$ at any time is described by the reduced density matrix $\rho_S(t)={\rm Tr}_{LR}(\rho(t))$, which is obtained by taking the partial trace over the reservoirs. 
Different notions of distance between the time-evolved state $\rho_S(t)$ and the stationary state $\rho_S(\infty)$ can be used to characterize the relaxation dynamics, and, consequently, to investigate the Mpemba effect. A comparative discussion of several commonly used quantum-state distances, including their respective advantages and limitations, can be found in Ref.~\cite{acm-25}. The most convenient choice of diagnostic may, however, depend on the physical setting under consideration. For example, for quenches from an initial homogeneous state, the relative entropy between $\rho_S(t)$ and $\rho_S(\infty)$ provides a particularly simple probe of the Mpemba effect, since it can be approximated by the difference between the corresponding entanglement entropies~\cite{arc-25}. One can therefore take advantage of the powerful analytical and numerical methods developed to study entanglement dynamics. This simplification, however, is no longer available for the inhomogeneous quenches considered in the present work.
 
Here, we employ the normalised Frobenius distance
between $\rho_S(t)$ and $\rho_S(\infty)$, introduced in Ref.~\cite{fagotti-rdm-frob},
\begin{equation}
	\mathcal{D}\left( \rho_S(t), \rho_S(\infty) \right) = \dfrac{\sqrt{\text{Tr}\left(\rho_S(t)^2 + \rho_S(\infty)^2 - 2 \rho_S(t)\rho_S(\infty)\right)}}{\sqrt{\text{Tr}\left(\rho_S(t)^2 + \rho_S(\infty)^2\right)}}.
	\label{eq:frob-norm-gen}
\end{equation}
Our motivation for considering the Frobenius distance is that it provides a tractable diagnostic in the present setting. In free-fermionic systems, it can be studied efficiently both numerically and analytically via GHD. Moreover, the Frobenius distance is not merely a computationally convenient theoretical quantity, as it has also been measured experimentally in Ref.~\cite{arc-25}.

While the Frobenius distance~\eqref{eq:frob-norm-gen} is generally difficult to evaluate for large systems, its computation simplifies considerably for the class of models considered here. In particular, the Hamiltonians governing the dynamics are quadratic in the fermionic creation and annihilation operators, and the initial states~\eqref{eq:xx_inital_state} of each part are fermionic Gaussian states. Since quadratic Hamiltonians preserve Gaussianity, the time-evolved state~\eqref{eq:evolved_state}, and consequently its reduced density matrix $\rho_S(t)$, remain Gaussian at all times. Both $\rho_S(t)$ and the stationary state $\rho_S(\infty)$ are therefore fully characterized by their two-point correlation matrices, which we denote by $\Gamma_S(t)$ and $\Gamma_S(\infty)$, respectively~\cite{peschel-03, pe-09}. In general, the two-point correlation matrix associated with $\rho_S(t)$ is defined as
\begin{equation}
	\begin{gathered}
		\Gamma_S = 2
		\begin{pmatrix}
			\mathbb{I}-C_S^T & F_S \\
			F_S^{\dagger} & C_S
		\end{pmatrix} - \mathbb{I}, 
	\end{gathered}
	\label{eq:corr-mat}
\end{equation}
where $C_S$ and $F_S$ denote the restriction to $S$ of  
\begin{equation}\label{eq:C_F_corr}
\left[C\right]_{ln} = {\rm Tr} (\rho(t) c^{\dagger}_l c_n)\hspace{8pt}\text{and}\hspace{8pt}\left[F\right]_{ln} = {\rm Tr}(\rho(t) c_l c_n).
\end{equation}
The correlation matrix of $\rho_S(\infty)$ is defined similarly, but replacing  $\rho_S(t)$ with $\rho_S(\infty)$ in Eq.~\eqref{eq:corr-mat}.
As shown in Ref.~\cite{fagotti-rdm-frob}, the Frobenius distance between two Gaussian states  can then be rewritten solely in terms of their correlation matrices as
\begin{equation}\label{eq:frob-norm-gaus}
\mathcal{D}\left( \rho_S(t), \rho_S(\infty) \right) = \left[ 1 - \dfrac{2\{ \Gamma_S(t), \Gamma_S(\infty) \}}{\{ \Gamma_S(t), \Gamma_S(t) \} + \{ \Gamma_{S}(\infty), \Gamma_{S} (\infty)\}}\right]^{\frac{1}{2}},
\end{equation}
where 
\begin{equation}\label{eq:tr_rho_det_gamma}	
\{ \Gamma, \Gamma' \}  = \text{Tr}\left( \rho\left[\Gamma\right] \rho\left[\Gamma'\right] \right)=\left(\text{det}\left| \dfrac{\mathbb{I} + \Gamma\Gamma'}{2} \right|\right)^{\frac{1}{2}}.
\end{equation}
This formula will be crucial in our analysis. Both the reduced density matrix $\rho_S(t)$ and its stationary value $\rho_S(\infty)$ have dimension $2^\ell\times 2^\ell$, but their two-point correlations matrices $\Gamma_S(t)$ and $\Gamma_S(\infty)$ have dimension $2\ell\times 2\ell$. This drastically reduces the complexity of computing numerically their distance. Moreover, Eq.~\eqref{eq:frob-norm-gaus} will be the starting point to derive an analytic formula for $\mathcal{D}(\rho_S(t), \rho_S(\infty))$ in the hydrodynamic regime.  

\section{Homogeneous quantum quench}\label{subsec:frobdist-qpp}

Before moving to the study of the tripartite setup described in Section~\ref{subsec:quench-setup-xx}, we first consider the case of a homogeneous quantum quench at zero temperature, which is the basic setting in which the quantum Mpemba effect has been explored in closed systems~\cite{amc-23}. The reasons are two-fold. To our knowledge, the normalised Frobenius distance has not been employed as a probe of the Mpemba effect in free systems (although it was used in the experiment in Ref.~\cite{Joshi-24}), and it is worth comparing it with other observables. Moreover, we use this setting to derive an exact expression for the Frobenius distance within the quasiparticle picture, which will be useful for the tripartite quench.

Here the full system, $L\cup S\cup R$, is prepared at $t=0$ in a pure state $\ket{\Psi_0}$ and then it evolves unitarily as
\begin{equation}\label{eq:hom_q_quench}
\ket{\Psi(t)}=e^{-itH}\ket{\Psi_0}.
\end{equation}
Following Ref.~\cite{makc-24}, we will take as initial state the ground state of the XY spin chain,
\begin{equation}\label{eq:Ham_XY}
H_0=-\frac{1}{2}\sum_{j=-\infty}^\infty(c_j^\dagger c_{j+1}+\gamma c_{j}^\dagger c_{j+1}^\dagger+{\rm h.c.}+2hc_j^\dagger c_j)
\end{equation}
and as post-quench Hamiltonian
\begin{equation}\label{eq:Ham_XX_0}
H=-\sum_{j=-\infty}^\infty \dfrac{1}{2} \left(c_j^\dagger c_{j+1}+\text{h.c.}\right)
\end{equation}
Since the Hamiltonian~\eqref{eq:Ham_XX_0} is translationally invariant and quadratic, it can be diagonalised in momentum space, with dispersion relation $\epsilon(k)=-\cos(k)$.

For this quench, the correlation matrix $\Gamma_S(t)$ of the reduced density matrix $\rho_S(t)$ can be written as a block Toeplitz matrix~\cite{makc-24, fc-08},
\begin{equation}\label{eq:corr_xy_2_xx_quench}
	[\Gamma_S(t)]_{jj'} = \int_{-\pi}^{\pi} \frac{dk}{2\pi}\,
		e^{-ik (j - j')} \,\mathcal{G}(k,t), 
		\qquad j,j' = 1,\ldots,\ell ,
\end{equation}
where 
\begin{equation}
		\mathcal{G}(k,t) =
		\begin{pmatrix}
			\cos \Delta_k 
			& -i e^{-2 i t \epsilon(k)} \sin \Delta_k \\[6pt]
			i e^{2 i t \epsilon(k)} \sin \Delta_k 
			& -\cos \Delta_k
		\end{pmatrix} ,
	\label{eq:symb_xy_2_xx_quench}
\end{equation}
and
\begin{equation}
	\cos \Delta_k = 
	\frac{h - \cos(k)}
	{\sqrt{(h - \cos(k))^{2} + \gamma^{2} \sin^{2} k}}, \hspace{8pt}\hspace{8pt}
	\sin \Delta_k = 
	\frac{-\gamma \sin(k)}
	{\sqrt{(h - \cos(k))^{2} + \gamma^{2} \sin^{2} k}} .
	\label{eq:cdk-sdk-xy}
\end{equation}
At long times, $\rho_S(t)$ tends to the generalised Gibbs ensemble~\cite{cef-11, ac-18}, 
\begin{equation}\label{eq:gge_xy_2_xx_quench}
\rho_S(\infty)=e^{-\sum_k \lambda_k d_k^\dagger d_k}/Z,
\end{equation}
where $d_k^\dagger$ and $d_k$ are the Fourier creation and annihilation operators that diagonalise the Hamiltonian~\eqref{eq:Ham_XX_0}. Since $d_k^\dagger d_k$ are integrals of motion of this Hamiltonian, the Lagrange multipliers $\lambda_k$ are obtained by imposing that
the initial density of occupied modes
\begin{equation}
{\rm Tr}(\rho(0) d_k^\dagger d_k)=\frac{1-\cos\Delta_k}{2}\equiv n(k)
\end{equation}
is conserved under the evolution and, consequently, ${\rm Tr}(\rho_S(\infty)d_k^\dagger d_k)= n(k)$. The two-point correlation matrix of this ensemble is of block Toeplitz form, as in Eq.~\eqref{eq:corr_xy_2_xx_quench}, but with symbol
\begin{equation}\label{eq:corr_gge_xy_2_xx_quench}
\mathcal{G}(k,t\to\infty) =
		\begin{pmatrix}
			\cos \Delta_k 
			& 0 \\[6pt]
			0 
			& -\cos \Delta_k
		\end{pmatrix}.
\end{equation}

Given that the initial state is translationally invariant, we can use the quasiparticle picture~\cite{pc-cardy-05, alba-pc-17-p1,c-20} to obtain an analytic prediction for the Frobenius distance between $\rho_S(t)$ and
the generalised Gibbs ensemble~\eqref{eq:gge_xy_2_xx_quench} that describes its long time limit.  According to it, pairs of entangled quasiparticles are created in the quench and propagate ballistically with opposite velocities $\pm v_k$, where $v_k=\partial_k\epsilon(k)$. In the hydrodynamic limit $\ell, t \to \infty$ with $\zeta = t/\ell$ fixed, the leading contribution to the traces appearing in the definition~\eqref{eq:frob-norm-gen} of the Frobenius distance is determined by the pairs of quasiparticles shared between $S$ and the rest of the system. In fact, it is well-known that in this quench ${\rm Tr}(\rho_S(t)^2)$ behaves in the hydrodynamic limit as~\cite{fc-08, ac-18, alba-pc-17-p1} 
\begin{equation}\label{eq:purity_xy_2_xx_quench}
\log {\rm Tr}(\rho_S(t)^2)\sim
\ell\int_{-\pi}^\pi \frac{dk}{2\pi}\min(2\zeta|v_k|, 1)h(n(k)), \quad 
h(\lambda)=\log\left(\frac{1+(2\lambda-1)^2}{2}\right).
\end{equation}
Throughout the paper, we use the symbol $\sim$ to indicate that the corresponding result is exact in the hydrodynamic limit.
This expression has a clear interpretation in terms of quasiparticles. The function $\min(2t|v_k|, \ell)$ counts the number of pairs of quasiparticles with momentum $k$ shared between $S$ and the rest of the system and $h(n(k))$ is their contribution to $\log {\rm Tr}(\rho_S(t)^2)$. In particular in the limit $\zeta\to\infty$, this expression gives the purity of the subsystem stationary state,
\begin{equation}\label{eq:large_t_pur_xy_2_xx}
\log {\rm Tr}(\rho_S(\infty)^2)\sim\ell\int_{-\pi}^\pi \frac{dk}{2\pi}h(n(k)).
\end{equation}
Equations~\eqref{eq:purity_xy_2_xx_quench} and~\eqref{eq:large_t_pur_xy_2_xx} are two of the terms necessary to compute the Frobenius distance~\eqref{eq:frob-norm-gen}. To the best of our knowledge, for the third one, ${\rm Tr}(\rho_S(t)\rho_S(\infty))$, there is no a quasiparticle prediction. 

To derive it, we need to determine the contribution of each quasiparticle pair, i.e. the analogous of $h(n(k))$. This is given by the difference between $\log{\rm Tr}(\rho_S(t)\rho_S(\infty))$ at $t=0$ and $t\to\infty$. We can obtain its large-$\ell$ behaviour at $t=0$ using the Widom-Szeg\"o theorem for determinants of block Toeplitz matrices~\cite{widom-74}. According to it, if $\Gamma[\mathcal{G}]$ and $\Gamma[\mathcal{G}']$ are $2\ell\times 2\ell$ block Toeplitz matrices with symbol $\mathcal{G}$ and $\mathcal{G}'$ respectively, such as~\eqref{eq:symb_xy_2_xx_quench} and~\eqref{eq:corr_gge_xy_2_xx_quench}, then (see Appendix E in Ref.~\cite{fa-sm-23}) 
\begin{equation}
	\log \text{det} \left(\mathbb{I} + \Gamma\left[\mathcal{G}\right]\Gamma\left[\mathcal{G}'\right] \right) \sim \ell\int_{-\pi}^{\pi}\dfrac{dk}{2\pi}\log\text{det}\left(\mathbb{I}+\mathcal{G}(k)\mathcal{G}'(k)\right),
	\label{eq:logdet2}
\end{equation}
for $\ell\gg1$. Recalling Eq.~\eqref{eq:tr_rho_det_gamma}, this result directly gives that
\begin{equation}\label{eq:tr_rho_S_0_infty}
	\log{\rm Tr}(\rho_S(0)\rho_S(\infty)) \sim \ell \int_{-\pi}^{\pi}\dfrac{dk}{2\pi}h_0(n(k)),\hspace{8pt} \quad h_0(\lambda) = \frac{1}{2}\log\left(\frac{1 + 3 (2\lambda-1)^2}{4} \right).
\end{equation}
Since at $t\to\infty$, $\log{\rm Tr}(\rho_S(t)\rho_S(\infty))$ must tend to Eq.~\eqref{eq:large_t_pur_xy_2_xx}, then the contribution of a pair of quasiparticles with momentum $k$ created in the quench is $h(k)-h_0(k)$. Thus, at time $t$, we only have to take into account the number of pairs with momentum $k$ shared between $S$ and the rest, 
\begin{equation}
	\log\text{Tr}\left[ \rho_S(t) \rho_S(\infty) \right] \sim \log\text{Tr}\left[ \rho_S(0) \rho_S(\infty) \right] + \ell \int_{-\pi}^{\pi} \dfrac{dk}{2\pi}\left[h(n(k))-h_0(n(k))\right]\text{min}\left(2\zeta|v_k|,1\right).	\label{eq:qpp_sample_term}
\end{equation}

\begin{figure}[t]
	\centering
	\includegraphics[width=0.6\textwidth]{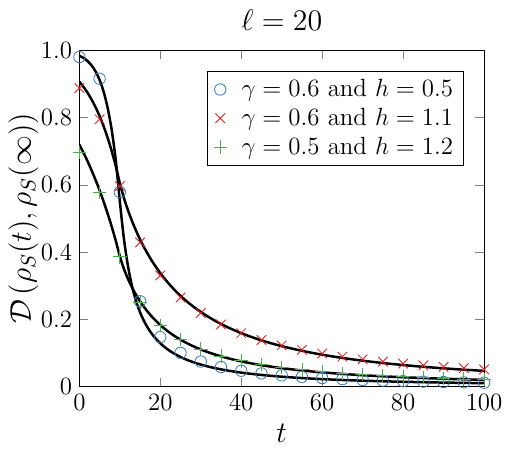}
	\caption{Time evolution of the Frobenius distance between the reduced density matrix of subsystem $S$, $\rho_S(t)$, and its stationary value, $\rho_S(\infty)$ in the homogeneous quench~\eqref{eq:hom_q_quench} to the XX spin chain from the ground state of the XY spin chain for different values of the parameters $h$ and $\gamma$. The symbols represent the result of exact numerics using Eq.~\eqref{eq:frob-norm-gaus} and the solid lines are the prediction~\eqref{eq:qpp_approx_xy_2_xx} of the quasiparticle picture.}
	\label{fig:qpp_xy_to_xx}
\end{figure}

Finally, combining the results in Eqs.~\eqref{eq:purity_xy_2_xx_quench}, \eqref{eq:large_t_pur_xy_2_xx}, and \eqref{eq:qpp_sample_term}, we conclude that in the hydrodynamic regime the Frobenius distance~\eqref{eq:frob-norm-gen} is given by
\begin{equation}
\mathcal{D}(\rho_S(t), \rho_S(\infty)) = \sqrt{1-2\dfrac{e^{\ell A(\zeta)+C(t, \ell)}}{e^{\ell A_P(\zeta)+C_P(t, \ell)} + e^{\ell A_{\infty}+C_\infty(\ell)}}},
\end{equation}
where the coefficients $A_p(\zeta)$,  $A_\infty$, and $A(\zeta)$ are the quasiparticle picture predictions reported in Eqs.~\eqref{eq:purity_xy_2_xx_quench},~\eqref{eq:large_t_pur_xy_2_xx}, and~\eqref{eq:qpp_sample_term}, respectively, while the $C$-terms are the corrections that vanish in the limit $\lim_{\ell\rightarrow\infty}C_{\cdot}(\ell,t)/\ell = 0$. For large $t$, since $\rho_S(t)$ approaches $\rho_{S}(\infty)$, it is reasonable to expect that these corrections are comparable and may largely cancel in the ratio above.
To test this hypothesis, we simply take all $C$-terms to be equal,
\begin{equation}\label{eq:qpp_approx_xy_2_xx}
\mathcal{D}(\rho_S(t), \rho_S(\infty)) \approx \sqrt{1-2\dfrac{e^{\ell A(\zeta)}}{e^{\ell A_P(\zeta)} + e^{\ell A_{\infty}}}},
\end{equation}
and examine how well this approximation
succeeds in reproducing the exact numerical results. 
The comparison is displayed in \cref{fig:qpp_xy_to_xx},
where the quasiparticle predictions (solid curves) exhibit excellent agreement with the numerics (symbols), obtained using Eq.~\eqref{eq:frob-norm-gaus}. Observe that the Frobenius distance tends to zero as expected, indicating the the subsystem $S$
relaxes to the stationary state~\eqref{eq:gge_xy_2_xx_quench} at long times. 

Another remarkable feature is that the Frobenius distance detects the occurrence of the quantum Mpemba effect.
Consider, for example, two initial states with parameters $(\gamma=0.6, h=0.5)$ and $(\gamma=0.6, h=1.1)$. The former is initially farther from the equilibrium state than the latter; however, as the system evolves, their Frobenius distances cross, and the state that was initially farther from equilibrium reaches it faster.
This phenomenon does not always occur, as is clear if we consider the initial states $(\gamma=0.6, h=1.1)$ and $(\gamma=0.5, h=1.2)$. In that case, the Frobenius distances do not cross, and the state initially farther from equilibrium approaches it more slowly.
The same has also been observed in this setup using different measures. First, via the entanglement asymmetry of $\rho_S(t)$, which monitors how the particle-number symmetry broken at $t=0$ is restored in $S$ at late times~\cite{makc-24}. Also with other measures of symmetry breaking such as the quantum Fisher information~\cite{yet-25} or the Gaussian asymmetry~\cite{tc-26}. It has also been probed via the relative entropy~\cite{arc-25} and the fidelity~\cite{pa-26} between $\rho_S(t)$ and $\rho_S(\infty)$. 

\subsection{Microscopic conditions for the Mpemba effect}

Let us derive the microscopic conditions under which the quantum Mpemba effect occurs in this quench, in terms of the Frobenius distance, and verify that they coincide with those previously obtained for other observables. We consider two initial states with parameters $(\gamma_1, h_1)$ and $(\gamma_2, h_2)$ such that their Frobenius distances satisfy
\begin{equation}
	\mathcal{D}(0,\gamma_1,h_1) < \mathcal{D}(0,\gamma_2,h_2),
	\label{eq:t0_mpemba_hom}
\end{equation}
that is, the initial state 1 is closer to the equilibrium state than the initial state 2. The quantum Mpemba effect occurs if and only if there exists a time $t_I$ such that, for all $t > t_I$, the ordering of the distances is reversed with respect to Eq.~\eqref{eq:t0_mpemba_hom},
\begin{equation}
	\mathcal{D}(t,\gamma_1,h_1) > \mathcal{D}(t,\gamma_2,h_2) \quad \forall t>t_I.
	\label{eq:tI_mpemba_hom}
\end{equation} 
Thus, the initial state 2 relaxes to the equilibrium state faster than state 1.

Let us first consider the condition at $t=0$, Eq.~\eqref{eq:t0_mpemba_hom}. Using the quasiparticle prediction in Eq.~\eqref{eq:qpp_approx_xy_2_xx}, it can be re-written as
\begin{equation}
	\left. e^{\ell \left( A_P(0) - A(0) \right) } + e^{\ell \left( A_\infty - A(0) \right) }{}\right|_{\gamma_1,h_1} < \left. e^{\ell \left( A_P(0) - A(0) \right) } + e^{\ell \left( A_\infty - A(0) \right) }{}\right|_{\gamma_2,h_2}.
\end{equation}
Using Eq.~\eqref{eq:purity_xy_2_xx_quench}, we have that $A_P(0)=0$ and the expression above simplifies as
\begin{equation}
\left. e^{-\ell A(0)}\left(1+e^{\ell A_\infty}\right)\right|_{\gamma_1,h_1}<\left. e^{-\ell A(0)}\left(1+e^{\ell A_\infty}\right)\right|_{\gamma_2,h_2}.
\end{equation}
%\begin{equation}
%	\left. e^{-\ell \int_{-\pi}^{\pi}\frac{dk}{2\pi}h_0(n(k))} \left[  1 + e^{\ell \int_{-\pi}^{\pi}\frac{dk}{2\pi}h(n(k))} \right] \right|_{\gamma_1,h_1} < \left. e^{-\ell \int_{-\pi}^{\pi}\frac{dk}{2\pi}h_0(n(k))} \left[  1 + e^{\ell \int_{-\pi}^{\pi}\frac{dk}{2\pi}h(n(k))} \right] \right|_{\gamma_2,h_2}
%	\label{eq:frob-hom-rewritten}
%\end{equation}
By noticing that $A_\infty<0$ for any $(\gamma, h)$, the terms in the brackets can be approximated to $1$ when $\ell\gg1$ and we obtain that the $t=0$ condition~\eqref{eq:t0_mpemba_hom} is equivalent to 
\begin{equation}
	\left. \int_{-\pi}^{\pi} \dfrac{dk}{2\pi} h_0(n(k)) \right|_{\gamma_1,h_1} > \left. \int_{-\pi}^{\pi} \dfrac{dk}{2\pi} h_0(n(k)) \right|_{\gamma_2,h_2}.
	\label{eq:t0_mpemba_hom_fin}
\end{equation}

Regarding the second condition~\eqref{eq:tI_mpemba_hom}, following the same steps as in Ref.~\cite{makc-24}, it can be expressed as a simple algebraic function of $(\gamma_1, h_1)$ and $(\gamma_2, h_2)$. First, \cref{eq:purity_xy_2_xx_quench} can be re-written as
\begin{equation}\label{eq:pur_Theta}
	\log \textrm{Tr}\left( \rho_{S}(t)^2 \right) \sim A_\infty - \ell \int_{-\pi}^{\pi} \dfrac{dk}{2 \pi}\left( 1 - 2 \zeta |v_k| \right) \Theta \left( 1 - 2 \zeta |v_k| \right) h(n(k)),
\end{equation}
where we used the identity
\begin{equation}
	1 - \textrm{min}\left( 2 \zeta |v_k| , 1 \right) = \left( 1 - 2 \zeta |v_k| \right) \Theta \left( 1 - 2 \zeta |v_k| \right),
\end{equation}
with $\Theta$ the Heaviside Theta function.
At large $\zeta$, the leading contribution in Eq.~\eqref{eq:pur_Theta} comes from the slowest modes, i.e. those around the modes with zero group velocity, $k=0$ and $k=\pi$. Expanding around these modes, Eq.~\eqref{eq:pur_Theta} behaves when $\zeta\to \infty$ as
\begin{equation}
	\log \textrm{Tr}\left( \rho_{S}(t)^2 \right) \sim A_\infty + \Upsilon_{\rm P}(h, \gamma) \dfrac{\ell}{\zeta^3},\quad \Upsilon_{\rm P}(h, \gamma)= \dfrac{1}{96 \pi} \dfrac{\gamma^2 \left(  h^2 + 1 \right)}{\left( h^2 - 1 \right)^2 }.
	\label{eq:purity_hom_asymptotic}
\end{equation}
Similarly, by recasting \cref{eq:qpp_sample_term}
\begin{equation}
	\log \textrm{Tr}\left( \rho_{S}(t)\rho_{S}(\infty) \right) \sim A_\infty - \ell \int_{-\pi}^{\pi} \dfrac{dk}{2 \pi}\left( 1 - 2 \zeta |v_k| \right) \Theta \left( 1 - 2 \zeta |v_k| \right) \left[ h(n(k)) - h_0(n(k)) \right],
\end{equation}
and expanding the second term around the slowest modes, $k=0$ and $\pi$, we get
\begin{equation}
	\log \textrm{Tr}\left( \rho_{S}(t)\rho_{S}(\infty) \right) \sim A_\infty +\Upsilon(h, \gamma)  \dfrac{\ell}{\zeta^3},\quad \Upsilon(h, \gamma)=\dfrac{1}{384 \pi} \dfrac{\gamma^2 \left(  h^2 + 1 \right)}{\left( h^2 - 1 \right)^2 }.
	\label{eq:cross_hom_asymptotic}
\end{equation}
Plugging \cref{eq:purity_hom_asymptotic,eq:cross_hom_asymptotic} into \cref{eq:tI_mpemba_hom}, we obtain upon squaring the later
\begin{equation}
	\dfrac{ 2 e^{\Upsilon(h_1, \gamma_1)\frac{\ell}{\zeta^3}}}{1 + e^{\Upsilon_{\rm P}(h_1, \gamma_1) \frac{\ell}{\zeta^3}}}  < \dfrac{ 2 e^{\Upsilon(h_2, \gamma_2) \frac{\ell}{\zeta^3}}}{1 + e^{\Upsilon_{\rm P}(h_2, \gamma_2) \frac{\ell}{\zeta^3}}},
\end{equation}
or, alternatively,
\begin{equation}
    e^{\Upsilon(h_1, \gamma_1)\frac{\ell}{\zeta^3}}\cosh\left( 2 \Upsilon(h_1, \gamma_1) \dfrac{\ell}{\zeta^3} \right) > e^{\Upsilon(h_2, \gamma_2)\frac{\ell}{\zeta^3}}\cosh\left( 2 \Upsilon(h_2, \gamma_2) \dfrac{\ell}{\zeta^3} \right).
\end{equation}
Since these functions are monotonic, the condition reduces to
\begin{equation}
\Upsilon(h_1, \gamma_1) > \Upsilon(h_2, \gamma_2),
\end{equation}
or, directly in terms of the parameters of the initial state,
to
\begin{equation}
	\dfrac{\gamma_1^2 \left(  h_1^2 + 1 \right)}{\left( h_1^2 - 1 \right)^2 } > \dfrac{\gamma_2^2 \left(  h_2^2 + 1 \right)}{\left( h_2^2 - 1 \right)^2 }.
    \label{eq:tinf_mpemba_hom_fin}
\end{equation}
Plugging in the different values of $\gamma$ and $h$ appearing in \cref{fig:qpp_xy_to_xx}, it is easy to see that the pairs of initial states satisfying \cref{eq:t0_mpemba_hom_fin,eq:tinf_mpemba_hom_fin} exhibit the quantum Mpemba effect, whereas those that do not satisfy \cref{eq:t0_mpemba_hom_fin,eq:tinf_mpemba_hom_fin} do not exhibit the Mpemba effect.

\section{XX spin chain}\label{sec:xx}

In this section, we study the tripartite quench in Fig.~\ref{fig:tripartition}, assuming that the subsystems are described by the Hamiltonian of a XX spin-$1/2$ chain with vanishing external magnetic field. Upon a Jordan–Wigner transformation, these correspond to those of the tight-binding model,
\begin{equation}
\begin{gathered}
    H_L =- \sum_{j=-\infty}^{-1}\dfrac{1}{2}\left(c^{\dagger}_j c_{j+1} + \text{ h.c.}\right), \quad H_S =- \sum_{j=1}^{\ell-1}\dfrac{1}{2}\left(c^{\dagger}_j c_{j+1} + \text{ h.c.}\right), \quad \\H_R =- \sum_{j=\ell+1}^{\infty}\dfrac{1}{2}\left(c^{\dagger}_j c_{j+1} + \text{ h.c.}\right)
\end{gathered}
    \label{eq:xx_inital_state}
\end{equation}
 
These Hamiltonians are diagonalised by sine modes, $\sin(kn)$, rather than plane waves, $e^{ikn}$, as a consequence of the open boundary conditions in each segment of the initial tripartite configuration. Using these eigenmodes, we can express the two‑point correlation functions~\eqref{eq:C_F_corr} of the initial state in the form
\begin{equation}
	C_{ln}(t=0) =
	\begin{cases} 
		\dfrac{2}{\ell+1}\displaystyle\sum_{k=1}^{\ell} \sin\left(\dfrac{\pi k}{\ell+1}l\right) \sin\left(\dfrac{\pi k}{\ell+1}n\right)
		\dfrac{1}{e^{\beta_s\epsilon\left(\frac{\pi k}{\ell+1}\right)}+1}, \, \, l,n\in S, \\
        \\
		\dfrac{2}{\pi}\displaystyle\int_{0}^{\pi}dk\dfrac{\sin(kl)\sin(kn)}{e^{\beta_b\epsilon(k)}+1}, 
        \quad l,n\in L\,\, \text{or}\,\, l, n\in R, \\
        \\
		0, \hspace{5pt}\text{ otherwise},
	\end{cases}.
	\label{eq:xx_corr_initial_state}
\end{equation}
and, since the Hamiltonians~\eqref{eq:xx_inital_state} commute with the particle number operator $Q=\sum_j c_j^\dagger c_j$,
\begin{equation}
 F_{ln}(t=0)=0.
\end{equation}
 We also observe that, for the baths, the correlation function at $t=0$ can be expressed in momentum space as
$$C_{ln}(t=0) =\int_{-\pi}^{\pi}\dfrac{dk}{2\pi}\dfrac{e^{ik(l-n)}}{e^{\beta_b\epsilon(k)}+1} - \int_{-\pi}^{\pi}\dfrac{dk}{2\pi}\dfrac{e^{ik(l+n)}}{e^{\beta_b\epsilon(k)}+1},\quad  l,n\in{L},\,\, \text{or}\,\, l,n\in{R},$$
which has the structure of a Toeplitz+Hankel matrix, due to the open boundary condition~\cite{fc-11}.

From $t=0$ onwards, we evolve the entire system under the homogeneous infinite XX Hamiltonian,
\begin{equation}
	H =- \sum_{j=-\infty}^{\infty} \dfrac{1}{2} \left(c^{\dagger}_j c_{j+1} + \text{ h.c.} \right)
	\label{eq:Hamiltonian_XX}
\end{equation}
This global Hamiltonian smoothly connects the baths $L$ and $R$ and the chain $S$, allowing the initially disconnected components to evolve as a single infinite system. 

As the initial state in \cref{eq:xx_inital_state} is Gaussian and the dynamics is governed by a quadratic Hamiltonian, the time evolution of the full system is completely determined by the propagation of the two‑point correlation functions. Following Ref.~\cite{eisler-peschel-07}, their time evolution is given by
\begin{equation}
	C_{jl}(t) = i^{\,l-j} \sum_{m,n\in \mathbb{Z}} i^{\,m-n} J_{j-m}(t) C_{mn}(0) J_{l-n}(t),
	\label{eq:xx_time_evol}
\end{equation}
 where $J_\alpha(z)$ denotes the Bessel function of first kind. The time evolution preserves the particle number symmetry and, consequently, $F_{ln}(t)=0$  for all $t$.

The only ingredient missing to calculate the Frobenius distance~\eqref{eq:frob-norm-gen} in this setting is the stationary state $\rho_S(\infty)$. We will determine it in the next subsection.

\subsection{Frobenius distance via GHD}
\label{subsec:frob-ghd-xx}

In the homogeneous quantum quench of   \cref{subsec:frobdist-qpp}, we found an analytic expression for the Frobenius distance using the quasiparticle picture. In the tripartite setting, this quasiparticle picture cannot be directly applied since the initial state is not translationally invariant. But we can rely on a generalisation to spatially inhomogeneous initially states. This framework was developed in Refs.~\cite{cdy-16,bcdf-16} and is now known as Generalised Hydrodynamics (GHD). 

The basic premise of GHD is that, when considering the thermodynamic limit, the system can be approximated by coarse-grained homogeneous cells, each encompassing many lattice sites. Then, replacing the site $j$ by a continuous variable $x$, the system can be locally described at large space-time scales by a quasiparticle momentum occupation function $n(k, x, t)$, whose time evolution is governed by a continuity equation. For the initial state in our tripartite setting, it corresponds to the Fermi-Dirac distribution with temperature $1/\beta_s$ in $S$ and $1/\beta_b$ in $L\cup R$,
\begin{equation}
	n(k,x,t=0) =
	\begin{cases} 
		\dfrac{1}{e^{\beta_b\epsilon(k)} + 1}, & x \in R\cup L, \\
		\dfrac{1}{e^{\beta_s\epsilon(k)} + 1}, & x \in S.
	\end{cases}
	\label{eq:xx_ghd_initial_state}
\end{equation}
For free systems, the time evolution of $n(k,x,t)$ in the hydrodynamic regime $x,t\to\infty$ with $x/t$ finite is precisely given by the Euler equation~\cite{bcdf-16, essler-22}
\begin{equation}
	\left(\partial_t + v_k\partial_x\right)n(k,x,t) = 0 \implies\hspace{4pt} n(k,x,t) = n(k,x-v_kt,0),
	\label{eq:xx_ghd_evol}
\end{equation}
where $v_k=\sin(k)$ is the group velocity of the Hamiltonian~\eqref{eq:Hamiltonian_XX}. Observe that, according to this result, starting from Eq.~\eqref{eq:xx_ghd_initial_state}, the local density of occupied modes tends, in the long-time limit, to
\begin{equation}\label{eq:ss_xx}
	n_{\textrm{ss}}(k) = n(k, x, t \to \infty) = \frac{1}{e^{\beta_b \epsilon(k)} + 1}
\end{equation}
for $x\in S$. This coincides with the occupation function of the Gibbs ensemble,
$\rho(\infty)=e^{-\beta_b H}/Z$. 
Hence, the subsystem 
$S$ thermalises to the same temperature as the baths 
$L$ and $R$. This result has a natural interpretation 
in terms of quasiparticles. According to 
Eq.~\eqref{eq:xx_ghd_evol}, the quasiparticles initially 
located in the region $S$ propagate ballistically 
either to the left bath $L$, if $v_k<0$, or to the 
right bath $R$, if $v_k>0$. At the same time, they are 
replaced in $S$ by quasiparticles coming from $L$ and 
$R$. At long times, all quasiparticles in $S$ originate 
from the baths and are described by the Fermi–Dirac 
distribution~\eqref{eq:ss_xx}.
The two-point correlation matrix of the subsystem's stationary state is, therefore,
\begin{equation}
[C_S(\infty)]_{ln}= \int_{-\pi}^\pi \frac{dk}{2\pi}
\frac{e^{ik(l-n)}}{e^{\beta_b\epsilon(k)}+1},\quad l,n\in S.
\end{equation}
Together with the correlation matrix of the time-evolved state~\eqref{eq:xx_time_evol}, this allows us to efficiently compute their Frobenius distance numerically using Eq.~\eqref{eq:frob-norm-gaus}. 

\begin{figure}[t]
	\centering
	\includegraphics[width=\textwidth]{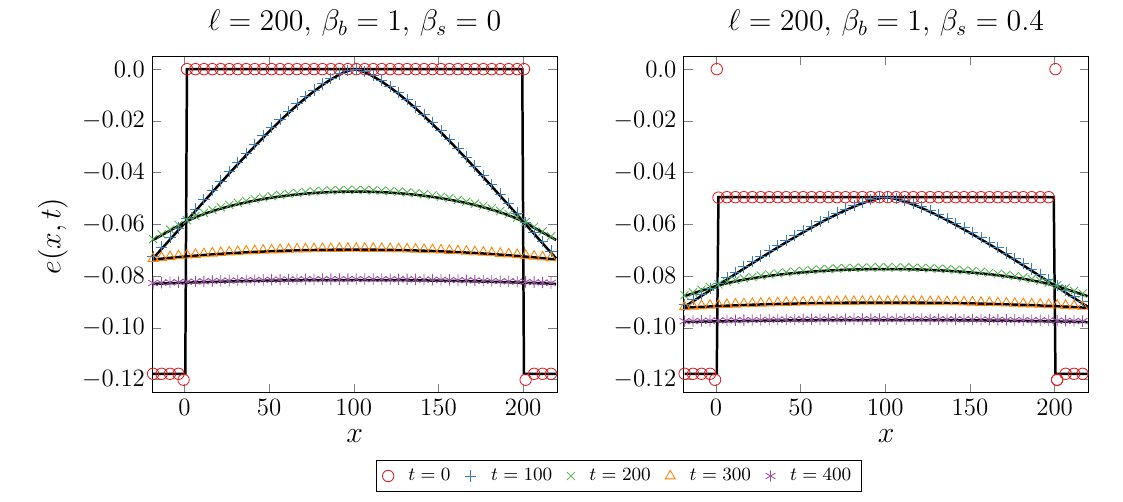}
	\caption{Energy density profile $e(x,t)$ in the tripartite quench of the XX spin chain at several times $t$. Symbols represent the exact numerical values, while solid lines are the GHD prediction~\eqref{eq:energy_density_GHD} obtained from the time-evolved local density of occupied modes~\eqref{eq:xx_ghd_evol}. In the left panel, the initial temperature of subsystem $S$ is $\beta_s=0$ (infinite temperature) and the semi-infinite subsystems 
    $L$ and $R$ are at temperature
    $\beta_b=1$. In the right panel, the same results are shown for $\beta_s=0.4$ and $\beta_b=1$.} 
	\label{fig:en-den-xx-prof}
\end{figure}

Before turning to the Frobenius distance, the local occupation density~\eqref{eq:xx_ghd_evol}  determines the evolution of local conserved charges, such as particle and energy densities, on the hydrodynamic scale. In our case, the particle density 
\begin{equation}
\varrho(x,t)=\int_{-\pi}^{\pi}\frac{dk}{2\pi}n(k, x, t)
\end{equation}
is not the most interesting observable in this case, because in the absence of an external magnetic field, particle-hole symmetry fixes $\varrho(x,t) = 1/2$ at any $x$ and $t$. We instead look at the local energy density
\begin{equation}\label{eq:energy_density_GHD}
    e(x,t) = \int\dfrac{dk}{2\pi}\epsilon(k)n(k,x,t),
\end{equation}
and compare it with Tr$\left( \rho(t) h_j \right)$, where $h_j = -\frac{1}{2}(c^{\dagger}_j c_{j+1} + c^{\dagger}_{+1} c_{i} )$, obtained numerically from Eq.~\eqref{eq:xx_time_evol}. We plot the result in \cref{fig:en-den-xx-prof} as a function of the of the site label $j$ at different times. We obtain an excellent agreement.

\begin{figure}[!t]
	\centering
	
	\begin{subfigure}{\textwidth}
		\centering
		\includegraphics[width=\textwidth]{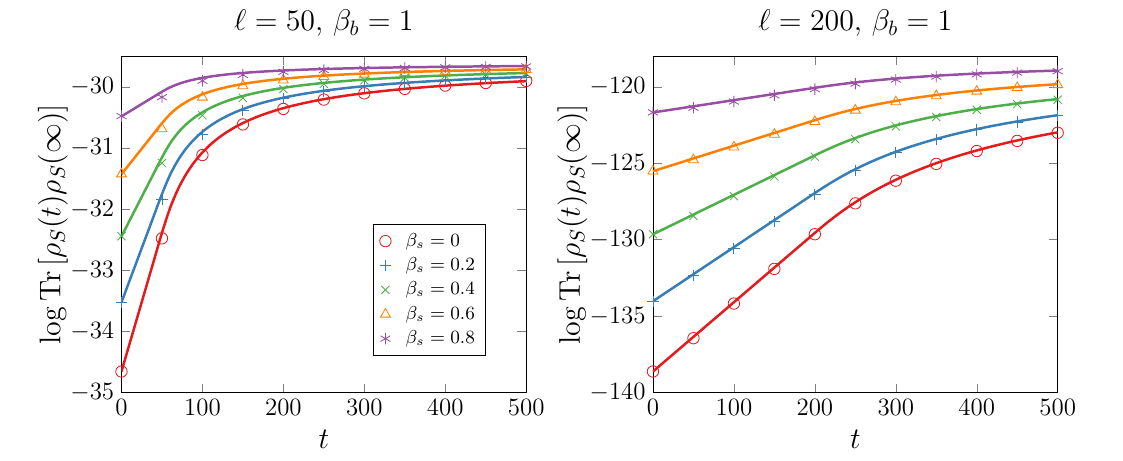}
		%\caption{}
		\label{fig:ghd-rtrgge-xx-b1}
	\end{subfigure}
	
	\medskip
	
	\begin{subfigure}{\textwidth}
		\centering
		\includegraphics[width=\textwidth]{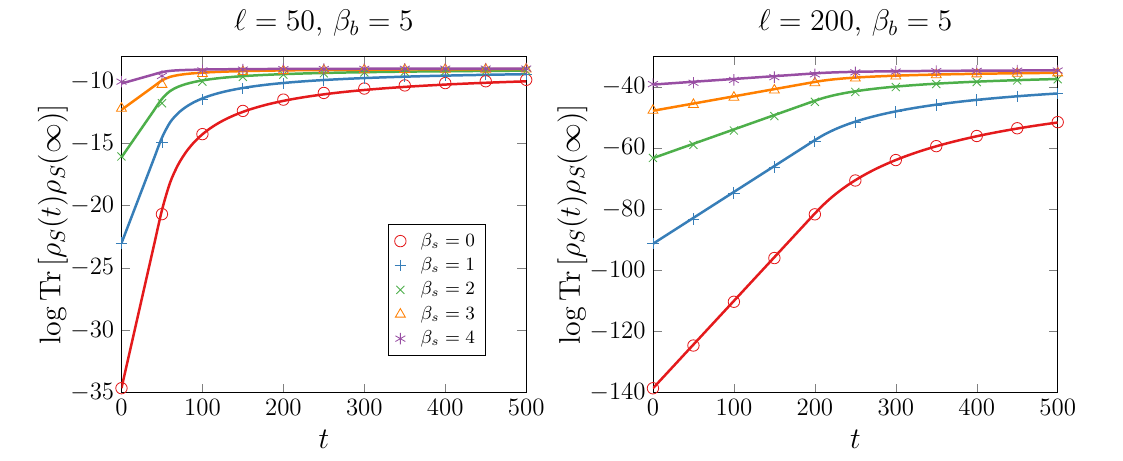}
		%\caption{}
		\label{fig:ghd-rtrgge-xx-b5}
	\end{subfigure}
	
	\caption{Time evolution of  $\log\text{Tr}(\rho_S(t)\rho_S(\infty))$ in the tripartite quench of the XX spin chain. Symbols denote the exact numerical value, calculated from the two-point correlation matrix~\eqref{eq:xx_time_evol}. Solid lines  are the GHD prediction in Eq.~\eqref{eq:cross_ghd_xx}.
 The semi-infinite subsystems $L$ and $R$ are prepared at temperature $\beta_b=1$ (upper panels) and $\beta_b=5$ (lower panels). We consider different initial temperatures $\beta_s$ for the subsystem $S$, together with two subsystem lengths, $\ell=50$ (left panels) and $\ell=200$ (right panels). Increasing $\ell$ leads to improved agreement between GHD prediction and the exact results.}
	\label{fig:ghd-rtrgge-xx}
\end{figure}

From the local occupation function~\eqref{eq:xx_ghd_evol}, we can further determine the leading-order behaviour, in the hydrodynamic limit, of $\log{\rm Tr}(\rho_S(t)^2)$ and $\log{\rm Tr}(\rho_S(t)\rho_S(\infty))$, which enter the definition of the Frobenius distance. This, in turn, requires the knowledge of the contribution of each mode to these two quantities. These contributions were derived in Eqs.~\eqref{eq:purity_xy_2_xx_quench} and~\eqref{eq:tr_rho_S_0_infty}, respectively, within the quasiparticle picture for the homogeneous quench, in terms of the density of occupied modes in the initial state $n(k)$. However, in that case, $\rho_S(t)$ breaks the particle-number symmetry, while $\rho_S(\infty)$ preserves it. Here, instead both respect particle number. We can deduce the mode contribution in this scenario by considering two homogeneous equilibrium Gaussian density matrices $\rho_1$ and $\rho_2$ with mode occupation densities $n_1(k)$ and $n_2(k)$ that preserve the particle-number symmetry. Their corresponding correlation matrices are
\begin{equation}
[\Gamma_j]_{nm}=\int_{-\pi}^\pi\frac{dk}{2\pi}e^{ik(n-m)}\mathcal{G}_j(k),\quad \mathcal{G}_j(k)=\left(\begin{array}{cc} 1-2n_j(k) & 0\\ 0 &2n_j(k)-1\end{array}\right).
\end{equation}
with $n, m=1, \dots, \ell$. The asymptotic behaviour of ${\rm Tr}(\rho_1\rho_2)$ for large $\ell$ is given by the Widom-Szeg\"o theorem~\eqref{eq:logdet2},
\begin{equation}
\log{\rm Tr}(\rho_1\rho_2)\sim \ell \int_{-\pi}^\pi \frac{dk}{2\pi}h'(n_1(k), n_2(k)), \quad 
h'(\lambda, \mu)=\log\left(\frac{ 1 + \left( 1 -2 \lambda \right) \left( 1 -2\mu\right)}{2} \right).
\end{equation}
Thus, we assume that $h'(n_1(k), n_2(k))$ is the contribution of the mode $k$ to $\log{\rm Tr}(\rho_1\rho_2)$. In our case, since we are joining three equilibrium states at $t=0$, we must evaluate the contribution of each mode at time $t$ using the local density $n(k,x,t)$, and then sum over all modes and over all points $x \in S = [0,\ell]$. As a result,
\begin{equation}\label{eq:purity_ghd_xx}
\log {\rm Tr}(\rho_S(t)^2)\sim
\int_S dx\int_{-\pi}^\pi \frac{dk}{2\pi}h(n(k, x, t)),
\end{equation}
and
\begin{equation}\label{eq:cross_ghd_xx}
	\log\text{Tr}\left[ \rho_S(t) \rho_S(\infty) \right] \sim \int_S dx \int_{-\pi}^{\pi}\dfrac{dk}{2\pi}h'(n(k, x, t), n(k, x, \infty)).
\end{equation}
In particular, Eq.~\eqref{eq:purity_ghd_xx} generalises to the tripartite quench the expression obtained in Ref.~\cite{bfpc-18} for the same quantity in the bipartite quench, where two semi-infinite chains at different temperatures are connected. Eq.~\eqref{eq:cross_ghd_xx} is instead a new result.
In Fig.~\ref{fig:ghd-rtrgge-xx}, we check this prediction (solid lines) against exact numerical results (symbols) for two choices of bath temperature, $\beta_b=1$ and $\beta_b=5$ and different subsystem $S$ initial temperatures. We obtain an excellent agreement that improves with the subsystem $S$ size.

\begin{figure}[t]
	\centering
	\begin{subfigure}{\textwidth}
		\includegraphics[width=\textwidth]{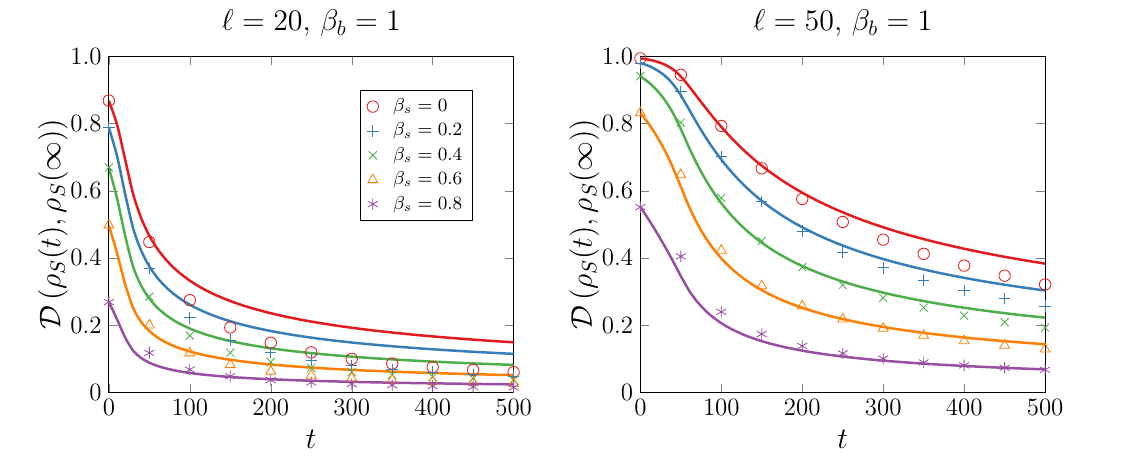}
		\label{fig:ghd-frob-xx-b1}
	\end{subfigure}
	\medskip
	\begin{subfigure}{\textwidth}
		\includegraphics[width=\textwidth]{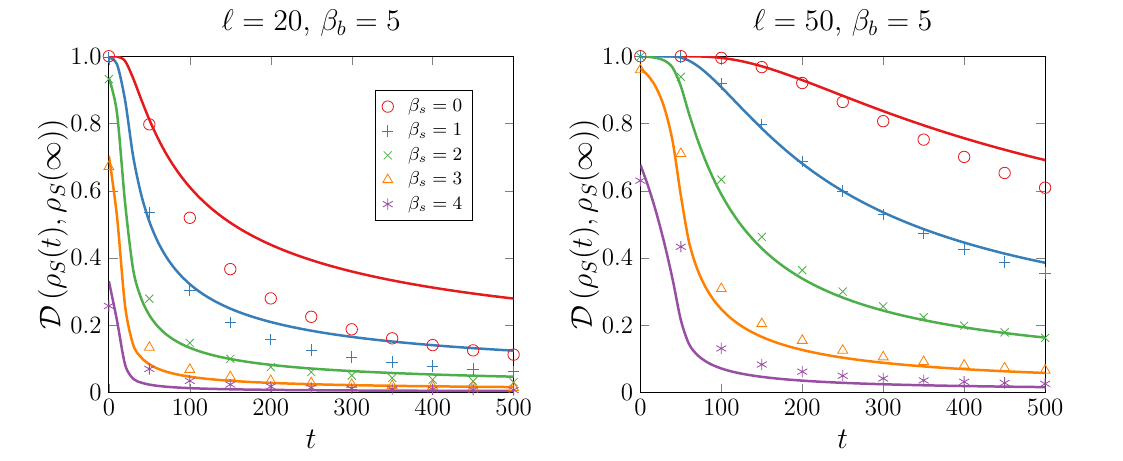}
		\label{fig:ghd-frob-xx-b5}
	\end{subfigure}
	\caption{Normalised Frobenius distance between the time-evolved state of $S$ and its stationary state in the tripartite quench in the XX spin chain. 
    Symbols are the exact value calculated through the two-point correlation matrices using Eq.~\eqref{eq:frob-norm-gaus}. Solid curves correspond to the GHD prediction~\eqref{eq:Frob_GHD}. The temperature of the baths $L$ and $R$ is $\beta_b=1$ (upper panels) and $\beta_b=5$ (lower panels). We take two different lengths for the subsystem $S$, $\ell=20$ (left panels) and $\ell=50$ (right panels).}
	\label{fig:ghd-frob-xx}
\end{figure}

Inserting the GHD expressions~\eqref{eq:purity_ghd_xx} and~\eqref{eq:cross_ghd_xx} in Eq.~\eqref{eq:frob-norm-gen}, we obtain the leading-order behaviour of the Frobenius distance between $\rho_S(t)$ and $\rho_S(\infty)$ in the hydrodynamic limit,
\begin{equation}\label{eq:Frob_GHD}
\mathcal{D}(\rho_S(t), \rho_S(\infty)) \approx \sqrt{1-2\dfrac{e^{A(\ell, t)}}{e^{A_P(\ell, t)} + e^{ A_{\infty}(\ell)}}},
\end{equation}
where $A_P(\ell, t)$ and $A(\ell, t)$ are given by Eqs.~\eqref{eq:purity_ghd_xx} and~\eqref{eq:cross_ghd_xx}, respectively, and 
\begin{equation}\label{eq:pur_XX_ss}
	 A_\infty(\ell) = \int_{S}dx \int_{-\pi}^{\pi}\dfrac{dk}{2\pi}\log\left[\frac{1 + \left(2 n_{\textrm{ss}}(k) - 1 \right)^2}{2}\right].
\end{equation}
is the stationary value 
of~\eqref{eq:purity_ghd_xx}. As in the homogeneous quench, we neglect the sub-leading corrections, which we expect to vanish in the large-$\ell$ limit.
In Fig.~\ref{fig:ghd-frob-xx}, we compare it against the exact time evolution. 
We again prepare the subsystem $S$ at different temperatures $1/\beta_s$, considering bath temperatures $\beta_b=1$ and $\beta_b=5$. As expected, the farther the initial temperature $1/\beta_s$ of subsystem $S$ is from that of the baths $1/\beta_b$, the larger the distance from the steady state. However, unlike in the homogeneous quench at zero temperature (cf. Fig.~\ref{fig:qpp_xy_to_xx}), the state that is farther from equilibrium now relaxes more slowly, and no Mpemba effect is observed in this case.

\subsection{Conditions for the Mpemba effect}

We can indeed use the GHD predictions to prove the absence of Mpemba effect in this setting. The necessary and sufficient conditions for the occurrence of the quantum Mpemba effect in this case are the following. If we prepare the subsystem $S$ at two different temperatures and $\beta_{s_1}$ $\beta_{s_2}$, for the same bath temperature $\beta_b$, such that
\begin{equation}
	\beta_{s1} < \beta_{s2} < \beta_{b} \Leftrightarrow \left.\mathcal{D}(t=0)\right|_{\beta_{s_1}}>\left.\mathcal{D}(t=0)\right|_{\beta_{s_2}}
\label{eq:t0_mpemba_inhom_xx}
\end{equation} 
then the Mpemba effect occurs when there exists a time $t_I$ for which
\begin{equation}
	\left. \mathcal{D}(t) \right|_{\beta_{s1}} <  \left. \mathcal{D}(t) \right|_{\beta_{s2}} \quad \forall t>t_I.
	\label{eq:tI_xx_inhom_mpemba}
\end{equation}

Let us now apply in this last condition the GHD formula~\eqref{eq:Frob_GHD}. The exponent $A_P(\ell, t)$, given by Eq.~\eqref{eq:purity_ghd_xx}, can be rewritten as
\begin{equation}
	A_P(\ell, t) = A_\infty(\ell) + \int_{S}dx \int_{-\pi}^{\pi}\dfrac{dk}{2\pi}\log\left[\dfrac{1 + \left(2 n(k,x,t) - 1 \right)^2}{1 + \left(2 n_{\textrm{ss}}(k) - 1 \right)^2}\right],
	\label{eq:purity_ghd_frob}
\end{equation}
where $A_\infty(\ell)$ is its stationary value~\eqref{eq:pur_XX_ss}. Similarly, Eq.~\eqref{eq:cross_ghd_xx} can be cast in the form
\begin{equation}
	A(\ell, t)= A_\infty(\ell) + \int_{S}dx \int_{-\pi}^{\pi}\dfrac{dk}{2\pi}\log\left[\dfrac{1 + \left(2 n(k,x,t) - 1 \right)\left(2 n_{\textrm{ss}}(k) - 1 \right)}{1 + \left(2 n_{\textrm{ss}}(k) - 1 \right)^2}\right].
	\label{eq:cross_ghd_frob}
\end{equation}
The crucial point is that, in \cref{eq:purity_ghd_frob,eq:cross_ghd_frob}, the modes for which $n(k,x,t) = n_{\textrm{ss}}(k)$, i.e., those that have thermalised at time $t$, give a vanishing contribution to the integral. Consequently, at long times the leading contribution to the integrals in \cref{eq:purity_ghd_frob,eq:cross_ghd_frob} comes from the modes that thermalise last. For $t$ large enough, all fluid cells away from the boundaries have thermalised except the one (say of width $\delta x$) around $x=\ell/2$. Also within this fluid cell, all modes except those with small group velocity, $|v_k|\ll 1$, have reached the stationary value, for $t$ large enough. In the XX chain, since $v_k = \sin (k)$, the slowest modes are those around $k^*=0$ and $\pi$, say those in $k\in \{k^*-\delta k/2,k^*+\delta k/2\}$, with $\delta k\ll 1$ . Then \cref{eq:purity_ghd_frob} in the limit $t\to\infty$ behaves as
\begin{equation}
	A_P(\ell, t) \sim A_\infty(\ell) + \dfrac{\delta x \delta k}{2\pi}\log\left[\dfrac{1 + \tanh^2(\beta_s\epsilon(0)/2)}{1+\tanh^2(\beta_b\epsilon(0)/2)}  
    \dfrac{1 + \tanh^2(\beta_s\epsilon(\pi)/2)}{1+\tanh^2(\beta_b\epsilon(\pi)/2)}
    \right].
\end{equation}
Since $\epsilon(k)=-\cos(k)$, this expression can be simplified to
\begin{equation}
    A_P(\ell, t) \sim A_\infty(\ell) +  \dfrac{\delta x\delta k}{\pi}\log\left[ \dfrac{1+\tanh^2(\beta_s/2)}{1+\tanh^2\left(\beta_b/2\right)} \right].
    \label{eq:purity_ghd_frob_asym}
\end{equation}
Applying the same reasoning in Eq.~\eqref{eq:cross_ghd_frob}, we find that at long times it behaves as
\begin{equation}
	A(\ell, t) \sim A_\infty(\ell) + \dfrac{\delta x\delta k}{\pi}\log\left[\dfrac{ 1+\tanh\left( \beta_s/2 \right) \tanh\left(\beta_b/2 \right) }{1 + \tanh^2\left(\beta_b/2\right)}\right].
	\label{eq:cross_ghd_frob_asym}
\end{equation}
Combining \cref{eq:purity_ghd_frob_asym,eq:cross_ghd_frob_asym} with~\eqref{eq:Frob_GHD} in the Mpemba condition~\eqref{eq:tI_xx_inhom_mpemba}, we obtain 
\begin{equation}
	\cosh{\left( \beta_{s1} \right)}\hspace{2pt}\textrm{sech}^2\left( \dfrac{\beta_{s1} + \beta_b }{2} \right) < \cosh{\left( \beta_{s2} \right)}\hspace{2pt}\textrm{sech}^2\left( \dfrac{\beta_{s2} + \beta_b }{2} \right).
    \label{eq:tinf_mpemba_inhom_xx}
\end{equation}
The function
\begin{equation}
    f(x)=\cosh{\left( x \right)}\hspace{2pt}\textrm{sech}^2\left( \dfrac{x + \beta_b }{2} \right),\quad 0<x<\beta_b,
    \label{eq:strictly_decreasing}
\end{equation}
satisfies
\begin{equation}
\dfrac{d}{dx}\log(f(x)) = \tanh(x) - \tanh\left(\dfrac{x+\beta_b}{2}\right). 
\end{equation}
When $0<x<\beta_b$, we have $x<\frac{x+\beta_b}{2}$ and, consequently,  $\tanh(x)<\tanh\left(\frac{x+\beta_b}{2}\right)$, since $\tanh$ is a strictly increasing function. From the latter inequality, it follows that
$$f'(x)< 0,\quad 0<x<\beta_b.$$
Hence $f(x)$ is a strictly decreasing function for $0<x<\beta_b$. The condition~\eqref{eq:tinf_mpemba_inhom_xx}, expressed in terms of $f(x)$, reads $f(\beta_{s1})<f(\beta_{s2})$. Since $f(x)$ is strictly decreasing in this regime, this condition requires $\beta_{s1} > \beta_{s2}$ in contradiction to \cref{eq:t0_mpemba_inhom_xx}. We therefore conclude that there is no Mpemba effect for this tripartite quench in the XX spin chain.

Although the direct Mpemba effect is absent in this setting, one may still ask whether the inverse Mpemba effect can occur~\cite{kcb-22}. This phenomenon arises when the subsystem $S$ is prepared at a lower temperature than the baths $R$ and $L$, such that the lower the initial temperature of $S$, the faster it reaches equilibrium. Thus, we consider that, at $t=0$,
\begin{equation}\label{eq:t0_inv_mpemba_cond}
\beta_b<\beta_{s2}<\beta_{s1}\Leftrightarrow 
\mathcal{D}(t=0)|_{\beta_{s1}}>\mathcal{D}(t=0)|_{\beta_{s2}},
\end{equation}
and suppose there is a time $t_I$ from which
\begin{equation}
	\left. \mathcal{D}(t) \right|_{\beta_{s1}} <  \left. \mathcal{D}(t) \right|_{\beta_{s2}} \quad \forall t>t_I.
	\label{eq:tI_xx_inhom_mpemba_inverse}
\end{equation}
Following the same steps as for the direct Mpemba effect, plugging the long-time GHD predictions~\eqref{eq:purity_ghd_frob_asym} and~\eqref{eq:cross_ghd_frob_asym} into the last condition yields Eq.~\eqref{eq:tinf_mpemba_inhom_xx} once again. In this case, since $x>\beta_b$, the function $f(x)$ in Eq.~\eqref{eq:strictly_decreasing} is strictly increasing for all $x>\beta_b$. Therefore, if the condition~\eqref{eq:tI_xx_inhom_mpemba_inverse} is satisfied, it follows that $\beta_{s1}<\beta_{s2}$, in contradiction with the $t=0$ condition in Eq.~\eqref{eq:t0_inv_mpemba_cond}. 
The inverse Mpemba effect is not possible too.

In both the homogeneous quench revisited in Sec.~\ref{subsec:frobdist-qpp} and in the present tripartite setting, relaxation can then be understood as the progressive depletion of the quasiparticle mode contributions, whit the late-time behaviour governed by the slowest modes. In the homogeneous quench, the Mpemba effect can arise because the modes that contribute more strongly to the distance from equilibrium also propagate faster and therefore relax earlier. Consequently, a state that is initially farther from equilibrium can eventually become closer to equilibrium than a state that was initially nearer. In the tripartite geometry, by contrast, our analysis shows that the mode occupations and their velocities are correlated in such a way that this competing mechanism is absent: at large times, the state that starts farther from equilibrium retains a larger contribution from the slowest modes throughout. As a result, the ordering of the relaxation curves is preserved at large times, and no Mpemba effect occurs.

\section{Transverse-field quantum Ising chain}
\label{sec:TFIC}

We now study the same tripartite quench introduced in \cref{subsec:quench-setup-xx}, but
considering another paradigmatic model: the transverse-field  Ising spin chain (TFIC). Thus, the Hamiltonian of each part before joining them are, after performing the Jordan-Wigner transformation, 
\begin{eqnarray}
	H_L &=& -\dfrac{1}{2} \sum_{j = -\infty}^{-1} \left[ c^{\dagger}_j c_{j+1} + c^{\dagger}_j c^{\dagger}_{j+1} + \mathrm{h.c.} \right]  + h \sum_{j = -\infty}^{0} c^{\dagger}_j c_j,
    \label{eq:Ising_JW_L}\\
    H_S &=& -\dfrac{1}{2} \sum_{j = 1}^{\ell-1} \left[ c^{\dagger}_j c_{j+1} + c^{\dagger}_j c^{\dagger}_{j+1} + \mathrm{h.c.} \right]  + h \sum_{j =1}^{\ell} c^{\dagger}_j c_j,
    \label{eq:Ising_JW_H_S}\\
    H_R &=& -\dfrac{1}{2} \sum_{j = \ell+1}^{\infty} \left[ c^{\dagger}_j c_{j+1} + c^{\dagger}_j c^{\dagger}_{j+1} + \mathrm{h.c.} \right]  + h \sum_{j =\ell+1}^{\infty} c^{\dagger}_j c_j,
	\label{eq:Ising_JW_R}
\end{eqnarray}
where $h$ is the intensity of the transverse magnetic field. For $t>0$, the full system evolves with the infinite TFIC Hamiltonian,
\begin{equation}
	H = -\dfrac{1}{2} \sum_{j = -\infty}^{\infty} \left[ c^{\dagger}_j c_{j+1} + c^{\dagger}_j c^{\dagger}_{j+1} + \mathrm{h.c.} \right]  + h \sum_{j = -\infty}^{\infty} c^{\dagger}_j c_j.
	\label{eq:Ising_JW_H}
\end{equation}
As in the XX spin chain, these Hamiltonians are quadratic in terms of fermionic creation and annihilation operators and, consequently, the total system density matrix is Gaussian. Thus, it suffices to compute the two-point correlation matrices. In contrast, the Hamiltonians now include pairing terms, $c_j^\dagger c_{j+1}^\dagger+c_j c_{j+1}$, which break particle-number conservation, leading to qualitative differences. The TFIC exhibits two phases: a ferromagnetic phase for $|h|<1$ and a paramagnetic phase for $|h|>1$.

\subsection{Correlation matrices}

The correlation matrix of the subsystem $S$, of size $\ell$, at $t=0$ is derived in detail in Appendix~\ref{app:obc}. Here we report the final expression of $C$ and $F$, defined in Eq.~\eqref{eq:C_F_corr},
\begin{equation}
	\left[C_S(0)\right]_{ln} = \dfrac{1}{2}\delta_{ln} - \dfrac{1}{4}\left( \left[Y\right]_{ln} + \left[Y\right]_{nl} \right), \quad \left[F_S(0)\right]_{ln} = \dfrac{1}{4}\left( -\left[Y\right]_{ln} + \left[Y\right]_{nl} \right), 
	\label{eq:corr_t_0_S_ising}
\end{equation}
where
\begin{equation}\label{eq:Y_Ising_OBC}
[Y]_{nl}=\sum_k \phi_{k,l} \psi_{k,n} \tanh\left( \dfrac{\beta \tilde{\epsilon}(k)}{2} \right),
\end{equation}
with $\phi_{k,l}$ and $\psi_{k,n}$ the eigenfunctions that diagonalise the finite-size Hamiltonian~\eqref{eq:Ising_JW_H_S}, see Eq.~\eqref{eq:basis2} in Appendix~\ref{app:obc}, and $\tilde{\varepsilon}(k)$ is the dispersion relation of the corresponding Bogoliubov modes, $\tilde{\varepsilon}(k)=\sqrt{1+h^2-2h\cos(k)}$. The sum in Eq.~\eqref{eq:Y_Ising_OBC} is over the momenta that satisfy the quantization condition
\begin{equation}
		\label{eq:quant_cond_2}
		\tan(Nk) = \dfrac{\sin(k)}{\frac{1}{h}-\cos(k)}.
\end{equation}
In Appendix~\ref{app:obc}, we also compute the two-point correlation matrices $C$ and $F$ in the semi-infinite baths $L$ and $R$ described by Eqs.~\eqref{eq:Ising_JW_L} and~\eqref{eq:Ising_JW_R}. They read
\begin{equation}
	\left[C_R(0)\right] = \dfrac{\delta_{ln}}{2} - \dfrac{S_{l-n} -  S_{l+n} - A_{l+n}}{2}, \quad \left[F_R(0)\right]_{ln} = \dfrac{A_{n-l}}{2} , \quad  l,n=\ell+1,\ell+2,\ldots.
\end{equation}
for subsystem $R$ and 
\begin{equation}
	\left[C_L(0)\right]_{l',n'} = \dfrac{\delta_{l',n'}}{2} - \dfrac{S_{l'-n'} -  S_{-l'-n'} -A_{-l'-n'} }{2}, \quad \left[F_L(0)\right]_{l',n'} = \dfrac{A_{n'-l'}}{2},  \quad  l',n'=\ldots,-2,-1
	\label{eq:cd-y3}
\end{equation}
for subsystem $L$. In both cases, 
\begin{equation}
		S_{l-n} = \displaystyle\int_{-\pi}^{\pi}\dfrac{dk}{2\pi}\cos\left( k(l-n) \right)\tanh\left(\dfrac{\beta_b\tilde{\epsilon}(k)}{2}\right)\dfrac{h-\cos(k)}{\tilde{\epsilon}(k)},
\end{equation}
and
\begin{equation}
		A_{l-n} =  -\displaystyle\int_{-\pi}^{\pi}\dfrac{dk}{2\pi}\sin\left( k(l-n) \right)\tanh\left(\dfrac{\beta_b\tilde{\epsilon}(k)}{2}\right)\dfrac{\sin(k)}{\tilde{\epsilon}(k)}.
\end{equation}

The time evolution of the full-system correlation matrix $\Gamma(t)$ under the infinite TFIC Hamiltonian~\eqref{eq:Ising_JW_H} is obtained in Appendix~\ref{app:ev_TIC}. Their entries evolve as 
\begin{equation}
		\Gamma_{nm}(t) = \sum_{l,j \in \mathbb{Z}} \mathbb{T}_{nl}(t) \Gamma_{lj}(0) \left[ \mathbb{T}^{\dagger}(t) \right]_{jm},
	\label{eq:Gamma_time_evol_2}
\end{equation}
where
\begin{equation}
\mathbb{T}(t) =
		\begin{pmatrix}
			U(t) & V(t) \\
			- \left[ V(t) \right]^{\dagger} & \left[ U(t) \right]^{\dagger}
		\end{pmatrix},
\end{equation}
and $U(t)$ and $V(t)$ are given by
\begin{equation}\label{eq:U_Wnm_Tevol}
	\left[ U(t) \right]_{nl}  = \int_{-\pi}^{\pi} \dfrac{dk}{2 \pi} \cos(k(n-l)) \left( \cos ( \tilde{\epsilon}(k) t )-i \sin ( \tilde{\epsilon}(k) t ) \cos \Delta_k  \right)
\end{equation}
and
\begin{equation}
	\left[ V(t) \right]_{nl} = \int_{-\pi}^{\pi} \dfrac{dk}{2 \pi} \sin(k(n-l)) \sin ( \tilde{\epsilon}(k) t ) \sin \Delta_k  
	\label{eq:V_Wnm_Tevol}.
\end{equation}
Here $\cos\Delta_k$ and $\sin\Delta_k$ are the Bogoliubov angles that diagonalise~\eqref{eq:Ising_JW_H}. Their explicit expressions are given in Eq.~\eqref{eq:cdk-sdk-xy}, taking $\gamma=1$.

In analogy with the Bessel-function kernels that appear in the evolution of the XX spin-chain correlation matrix~\eqref{eq:xx_time_evol}, the contributions to the sums in \cref{eq:Wnm_time_evol_2} decay exponentially with distance, allowing for a controlled truncation when calculating the Frobenius distance with Eq.~\eqref{eq:frob-norm-gaus}. It remains to determine the final stationary state of the subsystem $S$. As in the XX spin-chain case, we can deduce it using GHD arguments. With GHD, we will also derive an expression for the Frobenius distance in the hydrodynamic limit $\ell\to\infty$, $t\to\infty$ with $t/\ell$ finite.

\subsection{Evolution of Frobenius the distance}
\label{subsec:frob-ghd-ising}

\begin{figure}[t]
	\centering
	\includegraphics[width=\textwidth]{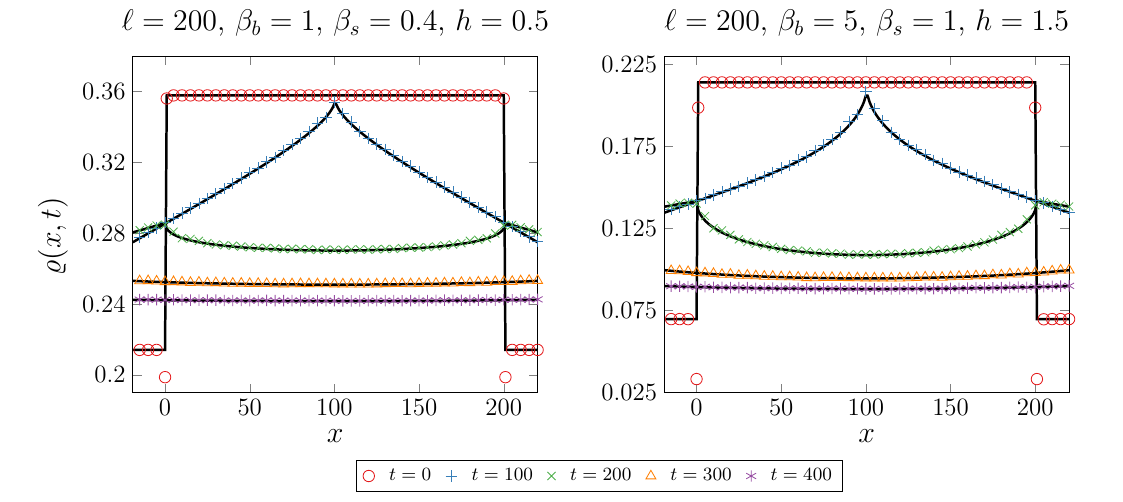}
	\caption{Particle density profile $\varrho(x,t)$ in the tripartite quench of the transverse-field Ising spin chain at several times $t$. Symbols represent the exact numerical values, while solid lines are the GHD prediction~\eqref{eq:dens_part_GHD_TIFC} obtained from the time-evolved local density of occupied modes~\eqref{eq:ising_ghd_evol}. In the left panel, the initial temperature of subsystem $S$ is $\beta_s=0.4$ and the semi-infinite subsystems 
    $L$ and $R$ are at temperature
    $\beta_b=1$. In the right panel, the same results are shown for $\beta_s=1$ and $\beta_b=5$. In both cases, we take a subsystem $S$ of length $\ell=200$.} 
	\label{fig:den-prof}
\end{figure}

Within the GHD framework introduced in Sec.~\ref{subsec:frob-ghd-xx}, the state of the system in the hydrodynamic regime is described by the function $n(k,x,t)$, which now denotes the local occupation density of the Bogoliubov fermionic modes that diagonalise the Hamiltonian~\eqref{eq:Ising_JW_H}. Since the subsystems are prepared at different finite temperatures, the initial configuration is specified by the corresponding Fermi–Dirac distributions,
\begin{equation}
	n(k,x,0) =
	\begin{cases} 
		\dfrac{1}{e^{\beta_b \tilde{\epsilon}(k)} + 1} &\forall\hspace{2pt} x \in R\cup L, \\
		\dfrac{1}{e^{\beta_s \tilde{\epsilon}(k)} + 1} &\forall\hspace{2pt} x \in S.
	\end{cases},
	\label{eq:ising_ghd_initial_state}
\end{equation}
Its time evolution is governed by the Euler equation
\begin{equation}
	\left(\partial_t + \tilde{v}(k)\partial_x\right)n(k,x,t) = 0 \implies\hspace{4pt} n(k,x,t) = n(k,x-\tilde{v}(k)t,0),
	\label{eq:ising_ghd_evol}
\end{equation}
where the group velocity is now given by the dispersion relation of the Bogoliubov modes, $\tilde{v}(k)=\partial_k \tilde{\epsilon}(k)$. 

As a first check, we can study the local particle density,
\begin{equation}\label{eq:dens_part_GHD_TIFC}
	\varrho(x,t)=\int_{-\pi}^{\pi}\dfrac{dk}{2\pi}n(k,x,t),
\end{equation}
which in this case exhibits a non-trivial evolution, in contrast to the XX spin chain. In Fig.~\ref{fig:den-prof}, we consider quenches for two values of the transverse field, $h=0.5$ and $h=1.5$,  with (inverse) bath temperatures $\beta_b = 1$ and $\beta_b = 5$ and the subsystem $S$ prepared at various temperatures $\beta_s$. Symbols are the exact local particle density ${\rm Tr}(\rho(t)c_j^\dagger c_j)$, calculated exactly from the diagonal entries of Eq.~\eqref{eq:Gamma_time_evol_2}, and solid lines are the GHD prediction~\eqref{eq:dens_part_GHD_TIFC}. One can also verify that the GHD prediction for the energy density agrees with the exact values, as we explicitly did for the XX spin chain, cf. Fig.~\ref{fig:en-den-xx-prof}.

\begin{figure}[!t]
	\centering
	
	\begin{subfigure}{\textwidth}
		\centering
		\includegraphics[width=\textwidth]{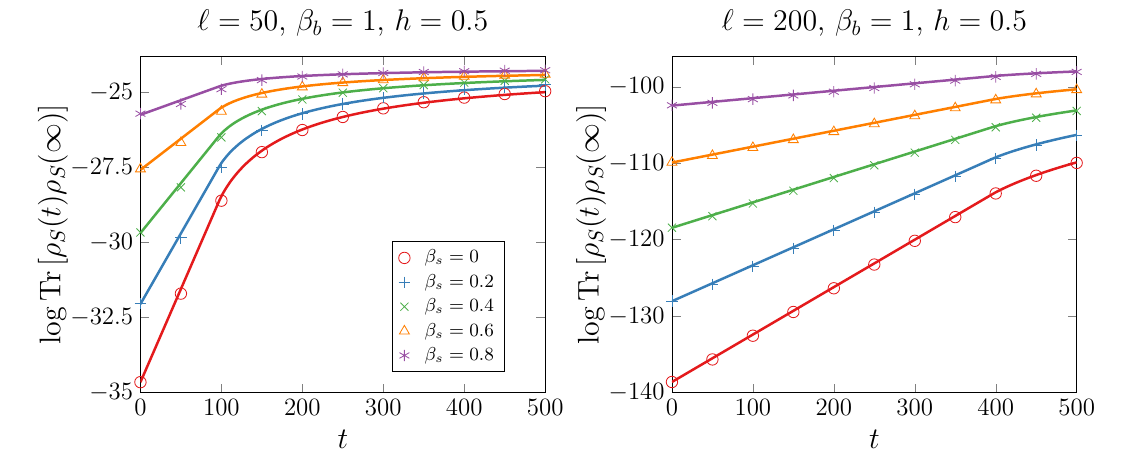}
		%\caption{}
		\label{fig:ghd-rtrgge-ising-h0p5-b1}
	\end{subfigure}
	
	\medskip
	
	\begin{subfigure}{\textwidth}
		\centering
		\includegraphics[width=\textwidth]{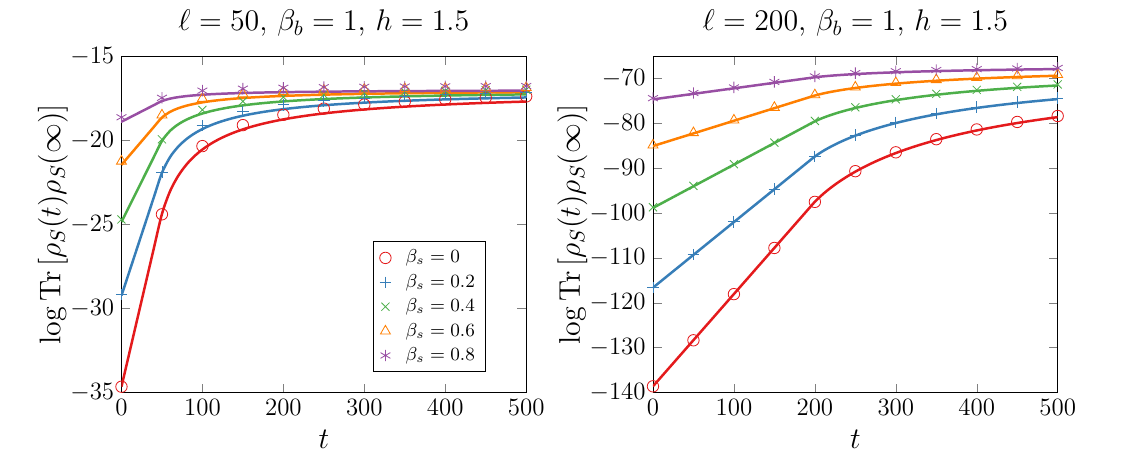}
		%\caption{}
		\label{fig:ghd-rtrgge-ising-h1p5-b1}
	\end{subfigure}
	
	\caption{Time evolution of  $\log\text{Tr}(\rho_s(t)\rho_s(\infty))$ in the tripartite quench of the quantum Ising spin chain  with $h=0.5$ (ferromagnetic phase, upper panels) and $h=1.5$ (paramagnetic phase, lower panels). Symbols denote the exact numerical value, calculated from the two-point correlation matrix~\eqref{eq:Gamma_time_evol_2}. Solid lines  are the GHD prediction in Eq.~\eqref{eq:cross_ghd_xx}, with $n(k,x,t)$ now the local density of Bogoliubov modes.
 The semi-infinite baths $L$ and $R$ are prepared at temperature $\beta_b=1$. We consider different initial subsystem temperatures $\beta_s$, for two sizes of subsystem $S$, $\ell=50$ (left panels) and $\ell=200$ (right panels). The agreement between GHD predictions and exact results improves as the subsystem size increases.}
	\label{fig:ghd-rtrgge-ising}
\end{figure}

Observe that, in the long time limit, the local occupation becomes
\begin{equation}
	n_{\textrm{ss}}(k) = n(k, x, t \to \infty) = \frac{1}{e^{\beta_b \tilde{\epsilon}(k)} + 1}
\end{equation}
for $x\in S$. This means that, as in the XX spin chain, the subsystem $S$ relaxes to the same state as the baths and 
the full system is eventually described by the Gibbs ensemble
$\rho(\infty)=Z^{-1}e^{-\beta H}$ with $H$ the infinite TFIC Hamiltonian~\eqref{eq:Ising_JW_H}. The two-point correlation matrix of this ensemble is 
\begin{equation}\label{eq:corr_gibbs_Ising}
[\Gamma_S(\infty)]_{jj'}=\int_{-\pi}^{\pi}\frac{dk}{2\pi}
e^{-ik (j - j')} \,\mathcal{G}_{\beta_b}(k)
\end{equation}
where 
\begin{equation}
		\mathcal{G}_{\beta_b}(k) =
		\tanh\left(\frac{\beta_b\tilde{\epsilon}(k)}{2}\right)\begin{pmatrix}
			\cos \Delta_k 
			& -i  \sin \Delta_k \\[6pt]
			i  \sin \Delta_k 
			& -\cos \Delta_k
		\end{pmatrix}.
	\label{eq:symb_xy_2_xx_quench}
\end{equation}
The functions $\cos\Delta_k$ and $\sin\Delta_k$ are given in Eq.~\eqref{eq:cdk-sdk-xy} with $\gamma=1$. 

We now have all the pieces to study the Frobenius distance between the subsystem state $\rho_S(t)$ under time evolution and its steady state $\rho_S(\infty)$. The corresponding correlation matrices are obtained from Eq.~\eqref{eq:Gamma_time_evol_2} and Eq.~\eqref{eq:corr_gibbs_Ising}, respectively, allowing us to compute the exact Frobenius distance numerically using Eq.~\eqref{eq:frob-norm-gaus}. We then compare these results with the GHD predictions derived in the previous section, in particular Eqs.~\eqref{eq:purity_ghd_xx} and \eqref{eq:cross_ghd_xx}, which can be straightforwardly applied to the Ising chain by inserting the corresponding local density of occupied Bogoliubov modes.

As a verification of this approach, in Fig.~\ref{fig:ghd-rtrgge-ising} we consider the term ${\rm Tr}(\rho_S(t)\rho_S(\infty))$ alone and confront the exact results (symbols), numerically obtained using Eq.~\eqref{eq:tr_rho_det_gamma}, with the GHD prediction conjectured in Eq.~\eqref{eq:cross_ghd_xx}, using now the local density $n(k,x,t)$ determined by Eqs.~\eqref{eq:ising_ghd_initial_state} and~\eqref{eq:ising_ghd_evol} (solid lines). We consider a chain in the paramagnetic regime ($h>1$) in the upper panels and a chain in the ferromagnetic regime ($0\leq h<1$) in the lower panels, for a subsystem $S$ prepared at different temperatures. In all cases, we find very good agreement, which improves with increasing subsystem size $\ell$.

\begin{figure}[t]
	\centering
	\begin{subfigure}{\textwidth}
		\includegraphics[width=\textwidth]{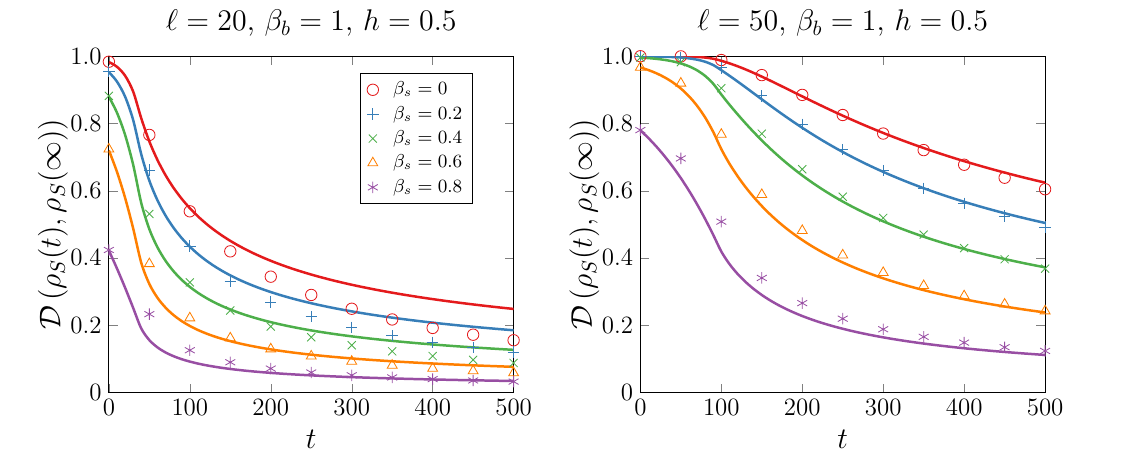}
		\label{fig:ghd-frob-ising-h0p5b1}
	\end{subfigure}
	\medskip
	\begin{subfigure}{\textwidth}
		\includegraphics[width=\textwidth]{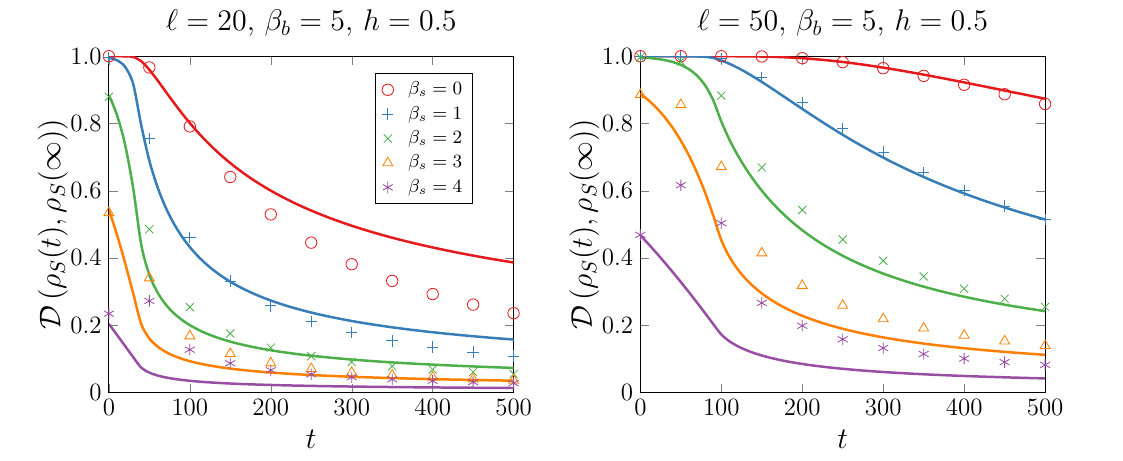}
		\label{fig:ghd-frob-ising-h0p5b5}
	\end{subfigure}
	\caption{Normalised Frobenius distance between the time-evolved state of $S$ and its stationary state in the tripartite quench in the Ising chain with magnetic field $h=0.5$ (ferromagnetic phase). 
    Symbols are the exact values calculated with Eq.~\eqref{eq:frob-norm-gaus}. Solid curves are the GHD prediction~\eqref{eq:Frob_GHD}, using the local density of Bogoliubov modes~\eqref{eq:ising_ghd_evol}. The temperature of the baths is $\beta_b=1$ (upper panels) and $\beta_b=5$ (lower panels). We consider two lengths for the subsystem $S$, $\ell=20$ (left panels) and $\ell=50$ (right panels).}
	\label{fig:ghd-frob-ising-h0p5}
\end{figure}

\begin{figure}[t]
	\centering
	\begin{subfigure}{\textwidth}
		\includegraphics[width=\textwidth]{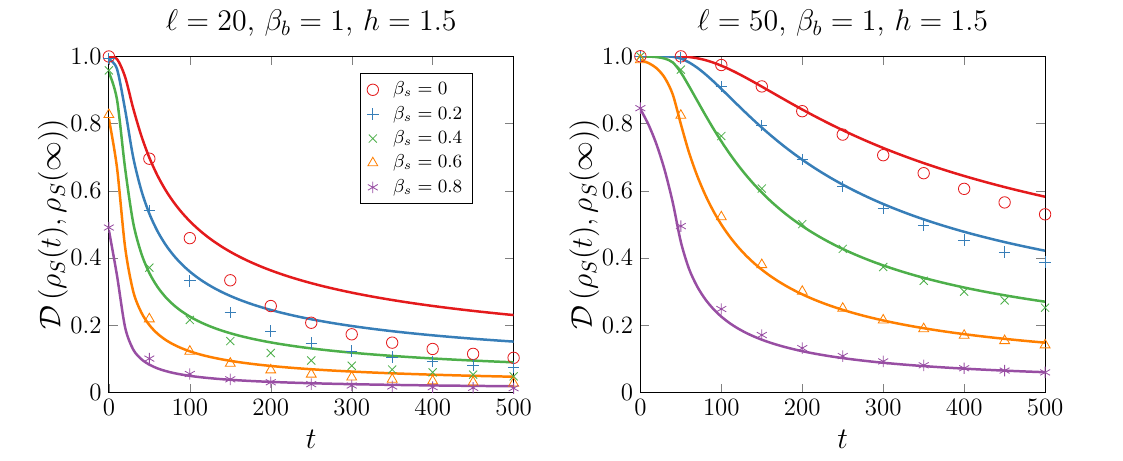}
		\label{fig:ghd-frob-ising-h1p5b1}
	\end{subfigure}
	\medskip
	\begin{subfigure}{\textwidth}
		\includegraphics[width=\textwidth]{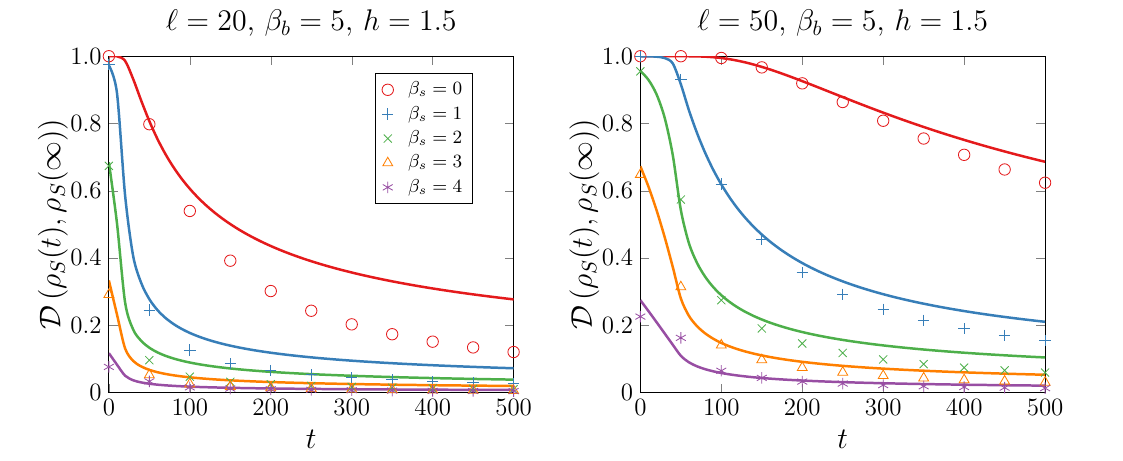}
		\label{fig:ghd-frob-ising-h1p5b5}
	\end{subfigure}
	\caption{Same as \cref{fig:ghd-frob-ising-h0p5}, but for transverse field $h=1.5$ (paramagnetic phase).}
	\label{fig:ghd-frob-ising-h1p5}
\end{figure}

In Figs.~\ref{fig:ghd-frob-ising-h0p5} and~\ref{fig:ghd-frob-ising-h1p5}, we show the results for the Frobenius 
distance between $\rho_S(t)$ and $\rho_S(\infty)$. The 
only difference between the figures is that in 
Fig.~\ref{fig:ghd-frob-ising-h0p5} we consider a chain 
in the ferromagnetic region while in 
Fig.~\eqref{fig:ghd-frob-ising-h1p5} the chain is in 
the paramagnetic regime. In the upper panels, the baths 
are prepared at $\beta_b=1$ while in the lower panels 
$\beta_b=5$. In the left panels, the subsystem size is 
$\ell=20$ and in the right ones is $\ell=50$. Overall, 
we obtain a good agreement between the exact Frobenius 
distance, calculated numerically using 
Eq.~\eqref{eq:frob-norm-gaus}, and the GHD prediction 
of Eq.~\eqref{eq:Frob_GHD}. As expected, the agreement 
between GHD and the exact numerics improves with 
increasing subsystem size $\ell$ as well as the baths 
temperature. This is evident in both figures: the $\ell = 50$ curves with $\beta_b = 1$ show better agreement with the GHD prediction than the $\ell = 20$ case and the $\ell = 50$, $\beta_b = 5$ case.

A particularly interesting feature appears in the ferromagnetic phase $|h|<1$ when both the subsystem $S$ and the baths $L$ and $R$ are prepared at low temperatures, $\beta_b, \beta_s \gg 1$. When $\beta_s$ is close to $\beta_b$, the normalised Frobenius distance exhibits a non‑monotonic behaviour: it initially increases before eventually decreasing, as shown in the lower panels of Fig.~\ref{fig:ghd-frob-ising-h0p5}. This non-monotonicity disappears when the subsystem is prepared at higher temperatures, as well as when the bath temperature is increased. It is also absent in the paramagnetic phase at all temperatures. A possible explanation for this non-monotonic behaviour is the presence of boundary bound states in the Ising chain with open boundary conditions in the ferromagnetic phase. The subsystem $S$ hosts these states at $t=0$ when $|h|<1$.  This feature is not captured by GHD, which always predicts a monotonous Frobenius distance. This effect is even more pronounced at $\beta_b=\beta_s=5$. In that case, the system begins in a tripartite product state with no correlations between the three regions, whereas the Gibbs ensemble at $\beta_b=5$ that describes the final equilibrium state is spatially homogeneous. One finds numerically a noticeable initial increase in $\mathcal{D}(t)$ before tending to zero, in contrast to GHD, which predicts $\mathcal{D}(t) = 0$ in that case. In general, boundary effects are not captured by GHD predictions, although their contribution becomes negligible when the temperatures of both the subsystem and the baths are increased.

In any case, as happened in the XX spin chain, the Mpemba effect is neither observed in the Ising chain. The larger is the difference of temperatures between the subsystem $S$ and the baths $L$ and $R$, the slower the subsystem thermalises. Again, we can use the GHD prediction~\eqref{eq:Frob_GHD} to prove the impossibility of Mpemba and inverse Mpemba effect in the Ising chain for any value of the external magnetic field. The necessary and sufficient conditions for the occurrence of the Mpemba effect in terms of the Frobenius distance are the same as in the XX spin chain. At $t=0$, as in \cref{eq:t0_mpemba_inhom_xx}, the subsystem $S$ is initialised at two different temperatures $\beta_{s_1} < \beta_{s_2}$, while the baths are prepared at the same temperature $\beta_b$ in both situations, and we demand that \cref{eq:tI_xx_inhom_mpemba} is satisfied. Again we can investigate whether this is possible by studying the latest modes that thermalise, i.e. the slowest modes at the middle of subsystem $S$, $x=\ell/2$.
The slowest modes are those around zero group velocity, $\tilde{v}(k_*)=0$, which for $|h|\neq 1$ correspond to $k^*=0, \pi$. Following the same steps as in the XX spin chain, we obtain that the Mpemba condition in \cref{eq:tI_xx_inhom_mpemba} implies
\begin{equation}
	\prod_{\{k_*\}} \cosh{\left( \beta_{s1} \left| \tilde{\epsilon}(k_*) \right| \right)}\hspace{2pt}\textrm{sech}^2\left( \dfrac{\beta_{s1} + \beta_b }{2} \left| \tilde{\epsilon}(k_*) \right| \right) < \prod_{\{k_*\}}  \cosh{\left( \beta_{s2} \left| \tilde{\epsilon}(k_*) \right|\right)}\hspace{2pt}\textrm{sech}^2\left( \dfrac{\beta_{s2} + \beta_b }{2} \left| \tilde{\epsilon}(k_*) \right| \right).
    \label{eq:tinf_mpemba_inhom_ising}
\end{equation}
These are just products of a rescaled version of the function $f(x)$ in \cref{eq:strictly_decreasing}. Hence, \cref{eq:tinf_mpemba_inhom_ising} implies $\beta_{s1}>\beta_{s2}$, in contradiction with \cref{eq:t0_mpemba_inhom_xx}. At the critical point $h= 1$, there is just one zero group velocity mode corresponding to $k_*=\pi$. which still obeys \cref{eq:tinf_mpemba_inhom_ising}. We therefore conclude that there is no Mpemba effect in the quantum Ising chain. This analysis can be further extended to the XY spin chain~\eqref{eq:Ham_XY}, yielding the same conclusion, namely that no Mpemba effect is present for the tripartite quench.

\section{Conclusions}\label{sec:conclusions}

We studied tripartite quenches in which a finite spin chain, initially prepared at a finite temperature, is suddenly connected to two semi-infinite chains at a lower temperature that act as thermal baths. The full system then evolves unitarily under a homogeneous Hamiltonian. This setup provides a quantum analogue of the original scenario underlying the classical Mpemba effect, where a system initially out of equilibrium is brought into contact with a colder environment.
We focused on two paradigmatic spin-$1/2$ models: the XX chain and the transverse-field Ising chain. The key distinction between them is that the former preserves a global $U(1)$ particle-number symmetry, whereas the latter does not. To monitor the relaxation dynamics of the subsystem, we employed the Frobenius distance from its stationary state as a quantitative measure of thermalisation.
Our analysis relies on two main technical ingredients. First, both models can be mapped onto free-fermion systems, implying that the state remains Gaussian throughout the evolution. As a consequence, the Frobenius distance can be computed efficiently and exactly in terms of two-point correlation matrices. Second, we exploit generalised hydrodynamics (GHD), which provides a systematic description of integrable systems evolving from spatially inhomogeneous initial conditions such as those considered here. Using GHD, we demonstrate that the subsystem relaxes to the same Gibbs ensemble as the baths and derive exact analytical expressions for the Frobenius distance in the hydrodynamic limit, thereby obtaining a complete characterisation of the thermalisation process.

We find that the Mpemba effect is absent in both the XX and transverse-field Ising chains. Instead, the time to relax increases monotonically with the temperature difference between the initial state of the finite chain and that of the baths: the farther the system is from equilibrium initially, the longer it takes to thermalise. Exploiting the analytical predictions of GHD, we establish the absence not only of the standard Mpemba effect, but also of its inverse counterpart, in which the baths are prepared at a higher temperature than the subsystem.
For completeness, we also revisit the homogeneous quantum quenches in free-fermion systems studied in Ref.~\cite{makc-24}, where the quantum Mpemba effect has previously been observed using a variety of probes~\cite{amc-23,pa-26,tc-26,yet-25,arc-25}, but not through the Frobenius distance. In this setting, we derive an exact expression for the Frobenius distance within the quasiparticle picture and use it to identify the microscopic conditions under which the Mpemba effect occurs. We find that these conditions coincide, in general, with those obtained from other probes, thereby confirming the robustness of the phenomenon and clarifying its physical origin.

Of course, our results do not rule out the possibility that the Mpemba effect may arise in tripartite quenches for other classes of models. One natural direction would be to consider free-fermion systems with long-range hopping and/or pairing terms, where qualitatively different transport properties may lead to novel relaxation phenomena. 
Another promising possibility is provided by interacting integrable models, such as the XXZ spin-$1/2$ chain~\cite{rvc-24}. 
In this case, the state is no longer Gaussian and the Frobenius distance cannot be reconstructed from two-point correlation functions alone. One must instead resort to more demanding numerical approaches, such as matrix-product-state techniques based on DMRG or TEBD. Nevertheless, the hydrodynamic description remains available: generalised hydrodynamics continues to provide an accurate large-scale description of the dynamics, although it must be supplemented with the appropriate Bethe-ansatz data~\cite{bcdf-16}. 
It would be extremely interesting to derive exact analytical expressions for the Frobenius distance by extending the space-time duality approach~\cite{bka-22,bcckr-22} to the tripartite quench considered here. Space-time duality has recently emerged as a powerful framework for obtaining exact results in non-equilibrium quantum dynamics, and only very recently it has been generalised to inhomogeneous settings~\cite{tckb-22}. However, its application to the tripartite geometry studied in this work remains an open problem. Solving it could provide a complementary and potentially exact description of the thermalisation dynamics, going beyond the hydrodynamic results for free theories presented here.

Finally, it would be interesting to extend our analysis to non-integrable systems, where the framework of GHD is no longer applicable. Promising candidates include the mixed-field Ising chain~\cite{bhore-25} and the long-range XX spin chain~\cite{ya-26,arc-25}, both of which have been shown to exhibit the quantum Mpemba effect when evolving from homogeneous initial states. Remarkably, evidence for this phenomenon has also been obtained experimentally~\cite{Joshi-24}. Investigating whether analogous effects arise in the thermalisation protocol considered here could provide valuable insight into the role played by integrability in suppressing or enhancing anomalous relaxation phenomena.
\newline

\textbf{Acknowledgments.} All authors acknowledge support from the European Research Council under the Advanced Grant no. 101199196 (MOSE).

\appendix

\begin{comment}
\section{Term wise comparison}
\label[appendix]{app:term-wise-comparison}
In this appendix, we compare the logarithms of the individual terms appearing in \cref{eq:frob-norm-gen} with their GHD predictions. Starting from subsystem size $\ell = 20$ and see that with increasing $\ell$, we observe that the agreement between GHD and the exact numerics steadily improves as $\ell$ increases. In \cref{fig:ghd-rtrgge-xx}, we show the comparison for $$\log\left(\text{Tr}\left(\rho_s(t)\rho_{\text{GGE},s}\right)\right)$$ 
and in \cref{fig:ghd-rtrt-xx}, the analogous quantity $$\log\left(\text{Tr}\left(\rho_s(t)\rho_s(t)\right)\right)$$
both evaluated at $\ell=20$ and $\ell=200$.

\begin{figure}[htbp]
	\centering
	
	\begin{subfigure}{\textwidth}
		\centering
		\includegraphics[width=\textwidth]{rtrt_b1.pdf}
		%\caption{}
		\label{fig:ghd-rtrt-xx-b1}
	\end{subfigure}
	
	\medskip
	
	\begin{subfigure}{\textwidth}
		\centering
		\includegraphics[width=\textwidth]{rtrt_b5.pdf}
		%\caption{}
		\label{fig:ghd-rtrt-xx-b5}
	\end{subfigure}
	
	\caption{Comparison of  $\log\left(\text{Tr}\left(\rho_s(t)\rho_s(t)\right)\right)$ obtained from GHD (solid lines) with exact numerical data (markers) for $\ell=20$ and $\ell=200$, in the transverse field ising chain. As in \cref{fig:ghd-rtrt-xx-b1}, the agreement improves with increasing subsystem size.}
	\label{fig:ghd-rtrt-xx}
\end{figure}

\end{comment}

\section{Spectrum and correlation matrix in the TFIC with OBC}
\label{app:obc}

In this Appendix, we review the spectrum and the two-point correlation matrix of the Gibbs ensemble of the transverse-field quantum Ising chain with open boundary conditions. This correlation matrix describes the state of subsystem $S$ at $t=0$ in the tripartite quench studied in Sec.~\ref{sec:TFIC}. The derivation closely follows Refs.~\cite{lsm-1961,cj-1987,fagotti-thesis,he-guo-17}, and we also implement the thermodynamic-limit treatment outlined in Ref.~\cite{fagotti-thesis}. For completeness, and to emphasise several points relevant to our application, we re-derive the essential results here.

\subsection{Spectrum}
The Hamiltonian of the TFIC can be rewritten in the generic form
\begin{equation}\label{eq:HXY_obc}
		H = \sum_{i,j} \left[ c^{\dagger}_i A_{ij} c_j + \dfrac{1}{2} \left( c^{\dagger}_i B_{ij} c^{\dagger}_j - c_i B_{ij} c_j \right) \right],
\end{equation}        
with
\begin{equation}
		A_{ij} = h \delta_{i,j} - \dfrac{1}{2} \left( \delta_{i+1,j} + \delta_{i,j+1} \right) = A_{ji} \textrm{ and } B_{ij} = -\dfrac{\gamma}{2} \left( \delta_{i+1,j} - \delta_{i,j+1} \right) = -B_{ji}.	
\end{equation}
To diagonalise this Hamiltonian, we introduce the fermionic operators
\begin{equation}
		\eta_k = \sum_k \left( g_{ki} c_i + h_{ki} c^{\dagger}_i \right), \quad
		\eta^{\dagger}_k = \sum_k \left( g_{ki} c^{\dagger}_i + h_{ki} c_i \right),
	\label{eq:bog_obc}
\end{equation}
where the coefficients $g_{ki}$ and $h_{ki}$ may be taken to be real. We require that they satisfy
\begin{equation}
	\left[ H, \eta_k \right] = - \lambda(k) \eta_k \Leftrightarrow \left[ H, \eta^{\dagger}_k \right] = \lambda(k) \eta^{\dagger}_k;
	\label{eq:Hobc_diag_cond}
\end{equation}
i.e., they diagonalise the Hamiltonian in \cref{eq:HXY_obc}. Inserting Eq.~\eqref{eq:bog_obc} into this condition, we obtain the set of constraints 
\begin{eqnarray}
		-\sum_i g_{ki} A_{ij} + \sum_i h_{ki} B_{ij} &= -\lambda(k) g_{kj}, \\
		-\sum_i g_{ki} B_{ij} + \sum_i h_{ki} A_{ij}& = -\lambda(k) h_{kj},
\end{eqnarray}
which can be re-cast in the form
\begin{eqnarray}	\label{eq:phipsi1}
		\sum_i \left( A - B \right)_{ji} \psi_{ki} &= \lambda(k)\phi_{kj}, \\\hspace{5pt} \sum_i \left( A + B \right)_{ji} \phi_{ki} &= \lambda(k)\psi_{kj},
\end{eqnarray}
where 
\begin{equation}
\phi_{ki} = g_{ki} + h_{ki} \text{ and } \psi_{ki} = g_{ki} - h_{ki}.
\end{equation}
To determine the functions $\phi_{ki}$ and $\psi_{ki}$, we can re-write \cref{eq:phipsi1} as the following eigenvalue equation,
\begin{eqnarray}
		\left[ \left( A + B \right) \left( A - B \right) \right]_{ln} \psi_{kn} &= \lambda(k)^2 \psi_{kl},\\
		\left[ \left( A - B \right) \left( A + B \right) \right]_{ln} \phi_{kn} &= \lambda(k)^2 \phi_{kl}.
	\label{eq:phipsi2}
\end{eqnarray}
We observe that 
\begin{equation}
\left( A + B \right) \left( A - B \right) = \tilde{\mathbb{I}} \left( A - B \right) \left( A + B \right)\tilde{\mathbb{I}},
\end{equation}
where $\tilde{\mathbb{I}}$  is the matrix containing ones along its anti‑diagonal and zeros elsewhere. This relation implies that it suffices to solve the eigenvalue problem for $\psi_{kn}$, after which the corresponding $\psi_{kl}$ are obtained simply as
\begin{equation}
	\phi_{kl}=\tilde{\mathbb{I}}_{ln}\psi_{kn}.
	\label{eq:phi-from-psi}
\end{equation}
Moreover, since $\left( A + B \right) \left( A - B \right)$ is real and symmetric, its eigenvalues $\lambda(k)^2$ are real, and the eigenvectors $\psi_{k,n}$, $\phi_{k,n}$ may be chosen to form a real orthonormal set:
$$\sum_k \psi_{k,n} \psi_{k,l} = \delta_{nl}  = \sum_k \phi_{k,n} \phi_{k,l}.$$

Let us now obtain the eigenmodes $\psi_{kn}$ and $\phi_{kn}$. A significant simplification occurs in our case, where the matrices $\left( A + B \right) \left( A - B \right)$ and $\left( A - B \right) \left( A + B \right)$ are tridiagonal and the eigenvalue problem \cref{eq:phipsi2} for the $\psi_{kl}$ modes reads
\begin{subequations}\label{eq:psi_eval}
	\begin{equation}\label{eq:psi_eval.1}
		\left( h^2 + 1 - \lambda^2(k) \right)\psi_{k,n} - h \left( \psi_{k,n-1} + \psi_{k,n+1}  \right)=0, \quad n \in \{ 1, \hspace{2pt}2, \hspace{1pt}\ldots, N-1 \},
	\end{equation}
	\begin{equation}\label{eq:psi_eval.2}
		\left( h^2 - \lambda^2(k) \right)\psi_{k,N} - h \psi_{k,N-1} = 0 .
	\end{equation}
\end{subequations}
In \cref{eq:psi_eval.1}, we impose the boundary condition $\psi_{k,0}=0$. As an ansatz for the eigenmodes, we take
\begin{equation}\label{eq:eig_psi}
\psi_{k,n} \propto \sin(n k),
\end{equation}
up to normalization. Plugging it into \cref{eq:psi_eval.1} yields 
\begin{equation}\label{eq:tfic_disp}
	\lambda^2(k) = 1 + h^2 -2h\cos(k) \implies \lambda(k) = \pm\tilde{\epsilon}(k),\quad \tilde{\epsilon}(k) = \sqrt{1 + h^2 -2h\cos(k)},
\end{equation}
while inserting the same ansatz into \cref{eq:psi_eval.2} leads to the quantisation condition for $k$,
\begin{equation}	
		\label{eq:quant_cond_1}
		h \sin\left( \left( N + 1 \right)k \right) = \sin\left( Nk \right), 
	\end{equation}
or, equivalently,
	\begin{equation}
		\label{eq:quant_cond_2}
		\tan(Nk) = \dfrac{\sin(k)}{\frac{1}{h}-\cos(k)}.
	\end{equation}
The latter expression is particularly convenient for determining the allowed values of $k$ numerically. If $k_{0}$ satisfies \cref{eq:quant_cond_1}-\eqref{eq:quant_cond_2}, then so does $-k_{0}$; in what follows, we select the convention $k_{0} > 0$. 

As noted earlier in \cref{eq:phi-from-psi}, if the eigenmodes $\psi_{kn}$ are of the form~\eqref{eq:eig_psi}, then the corresponding eigenmodes $\phi_{k,n}$ are given by
\begin{equation}
\phi_{k, n}\propto\sin\left( \left(N + 1 -n\right)k \right).
\end{equation}
However, it will be useful to rewrite this expression in a form more amenable to taking the thermodynamic limit. To this end, we first observe that
 \begin{equation}
 	\left( \dfrac{h\sin(k)}{\sin(Nk)} \right)^2 = \left( \dfrac{\sin(k)}{\sin(\left( N + 1 \right)k)} \right)^2 = \lambda(k)^2,
 	\label{eq:quant_energy_rel}
 \end{equation}
a relation that follows directly from \cref{eq:quant_cond_1}-\eqref{eq:quant_cond_2}. Using the connection between the quantisation condition and the single‑particle energies in \cref{eq:quant_energy_rel}, we arrive at
\begin{equation}
	h\sin(nk) - \sin\left((n-1)k\right) = \pm \lambda(k)\sin\left( \left(N + 1 -n\right)k \right),
	\label{eq:helper_thermodyanmic_limit}
\end{equation}
which removes the explicit $N$ dependence from the argument of the trigonometric function and thereby aids in taking the thermodynamic limit. 
One can numerically verify the following:
\begin{enumerate}
	\item \textbf{First choice of basis}
	\begin{equation}
		\psi_{k,n} = N_k \sinh(nk), \quad \phi_{k,n} = N_k \sin\left( \left(N + 1 -n\right)k \right)
		\label{eq:basis1}
	\end{equation}
	satisfies \cref{eq:phipsi1}, with the corresponding $\lambda(k)$ being positive for some values of $k$ and negative for other.
	\item \textbf{Second choice of basis}
	\begin{equation}
		\psi_{k,n} = N_k \sin(nk), \quad \phi_{k,n} = \left( N_k/\tilde{\epsilon}(k) \right) \left( h\sin(nk) - \sin\left((n-1)k\right) \right)
		\label{eq:basis2}
	\end{equation}
	also satisfies \cref{eq:phipsi1}, but with $\lambda(k) = \tilde{\epsilon}(k) $ for all values of $k$ (i.e., $\lambda(k)$ is positive for all values of $k$).
\end{enumerate}
Both bases are therefore consistent, provided the sign of $\lambda(k)$ is taken into account. Certain parts of the analysis are more conveniently carried out using the former basis, while others benefit from the latter. In both cases, the normalisation constant satisfies
\begin{equation}
	N^{-2}_k = \sum_{n=1}^{N} \sin^2(nk) = \dfrac{1}{2}\left( N - \dfrac{\cos\left( (N + 1)k \right)\sin(Nk)}{\sin(k)} \right) = \dfrac{N}{2} \left[ 1 - h\left( \dfrac{\cos(k) - h}{N \epsilon^2(k)} \right) \right],
	\label{eq:modes_norm}
\end{equation}
where the final expression follows from \cref{eq:quant_energy_rel}. 

\subsection{Correlation matrices}

To obtain the correlation matrices $C$ and $F$ in Eq.~\eqref{eq:C_F_corr} for the Gibbs ensemble of the TFIC with OBC, it is convenient to introduce the Majorana operators 
\begin{equation}
	a^x_j = c^{\dagger}_j + c_j, \quad a^y_j = i ( c_j - c^{\dagger}_j).
\end{equation}
In terms of them, 
\begin{equation}
	\left[C\right]_{ln} = \dfrac{1}{2}\delta_{ln} - \dfrac{1}{4}\left( \left[Y\right]_{ln} + \left[Y\right]_{nl} \right), \quad \left[F\right]_{ln} = \dfrac{1}{4}\left( -\left[Y\right]_{ln} + \left[Y\right]_{nl} \right), 
	\label{eq:cd-y}
\end{equation}
where
\begin{equation}
\left[Y\right]_{ln} := i{\rm Tr}(\rho a^x_l a^y_n).
\end{equation}
The Bogoliubov operators $\eta_k,\eta_k^{\dagger}$ can be expressed using the Majorana operators as
\begin{equation}
		\eta_k = \dfrac{1}{2}\sum_l \left( \phi_{kl} a_l^x - i \psi_{kl} a_l^y \right),\quad
		\eta^{\dagger}_k = \dfrac{1}{2}\sum_l \left( \phi_{kl} a_l^x + i \psi_{kl} a_l^y \right).
\end{equation}
Inverting these expressions,
\begin{equation}
		a_{l}^x=\sum_k \phi_{k,l}\left( \eta_k + \eta_k^{\dagger} \right),\quad  a_l^y = i\sum_k \psi_{k,l}\left( \eta_k - \eta_k^{\dagger} \right).
	\label{eq:Majarona-eta}
\end{equation}
In the Bogoliubov basis $\{ \eta_{k},\eta_{k}^\dagger \}$, the Hamiltonian takes the diagonal form
\begin{equation}
H=\sum_k\lambda_k \eta_{k}^\dagger\eta_{k}.
\end{equation}
Thus, using \cref{eq:Majarona-eta} together with the thermal expectation values
\begin{equation}
{\rm Tr}(\rho\eta^\dagger_k \eta_{k'}) = \delta_{kk'}\dfrac{1}{1+e^{\beta \lambda(k)}},\quad {\rm Tr}(\rho \eta_k \eta_{k'}) = 0,
\end{equation}
the two‑point correlation functions of the Majorana operators can be computed explicitly. After a straightforward algebra, we find
\begin{equation}
		{\rm Tr}(\rho a_l^x a_n^x) = \delta_{nl} = {\rm Tr}(\rho a_l^y a_n^y)  \hspace{8pt}\left( \text{basis independent} \right),
\end{equation}
and
\begin{equation}\label{eq:obc_ising_corr_0}
		i {\rm Tr}(\rho a_l^x a_n^y ) = \sum_k \phi_{k,l} \psi_{k,n} \tanh\left( \dfrac{\beta \lambda(k)}{2} \right) \hspace{8pt}\left( \text{in the basis of \cref{eq:basis1}} \right),
\end{equation}
or
\begin{equation}
		i {\rm Tr}(\rho a_l^x a_n^y ) = \sum_k \phi_{k,l} \psi_{k,n} \tanh\left( \dfrac{\beta \tilde{\epsilon}(k)}{2} \right) \hspace{8pt}\left( \text{in the basis of \cref{eq:basis2}} \right).
	\label{eq:obc_ising_corr}
\end{equation}
Inserting these expressions in Eq.~\eqref{eq:cd-y} gives the full correlation matrix in terms of the Dirac operators $c_j$ and $c_j^\dagger$. Once the correlation matrix has been assembled in this way, the non-Gaussianity can be computed using \cref{eq:frob-norm-gaus}.

We remark that it is essential to keep track of the sign of $\lambda(k)$ if we use \cref{eq:obc_ising_corr_0}. Indeed, in the zero‑temperature limit, we have
\begin{equation}
\tanh\left( \dfrac{\beta \lambda(k)}{2} \right) \xrightarrow{\beta\rightarrow\infty}\text{sgn}\left( \lambda(k) \right),
\end{equation}
which implies
\begin{equation}
i {\rm Tr}(\rho a_l^x a_n^y) \xrightarrow{\beta\rightarrow\infty} \sum_k \phi_{k,l} \psi_{k,n} \text{sgn}\left( \lambda(k) \right) \hspace{8pt}\left( \text{in the basis of \cref{eq:basis1}} \right).
\end{equation}
If, within the basis of \cref{eq:basis1}, one incorrectly assumes that $\lambda(k)>0 \hspace{2pt} \forall \hspace{2pt} k$, then the orthogonality of that basis leads to
\begin{equation}
i {\rm Tr}(\rho a_l^x a_n^y)\xrightarrow{\beta\rightarrow\infty}  \delta_{l+n,N},
\end{equation}
which in turn would imply
\begin{equation}
{\rm Tr}(\rho c_l c_n) \xrightarrow{\beta\rightarrow\infty}  0.
\end{equation}
However, this conclusion is erroneous: the Hamiltonian~\eqref{eq:HXY_obc} breaks the $U(1)$ particle-number symmetry, and therefore the correlations ${\rm Tr}(\rho c_l c_n)$ need not vanish, including in the zero-temperature limit $\beta\to\infty$. This discrepancy highlights the importance of correctly incorporating the sign of the quasiparticle energies $\lambda(k)$ when constructing the Majorana correlation functions in the basis of \cref{eq:basis1}.
 
\subsection{Bound states and the thermodynamic limit} For $|h|<1$, the spectrum of the open‑boundary TFIC contains a bound state. When $0<h<1$, this mode corresponds to a purely imaginary momentum $k = i \kappa$, where $\kappa>0$ satisfies, cf. Eq.~\eqref{eq:quant_cond_2},
\begin{equation}
	\tanh(N\kappa) = \dfrac{\sinh(\kappa)}{\frac{1}{h}-\cosh(\kappa)}.
	\label{eq:bound_state_k}
\end{equation}
For $-1<h<0$, this bound state corresponds to the mode $k = \pi + i \kappa$. In the thermodynamic limit $N\to\infty$, keeping $\kappa$ finite, \cref{eq:bound_state_k} reduces to
\begin{equation}
	\kappa = -\log h + \mathcal{O}\left( h^{2N} \right).
	\label{eq:bound_state_k_Ninf}
\end{equation}

We now examine the contribution of this bound state to the correlation functions $i{\rm Tr}(\rho a_l^x a_n^y)$ when $N\gg 1$. Using the basis in \cref{eq:basis1}, the mode functions $\psi_{k, n}$ and $\phi_{k, n}$  take the form for $k=i\kappa$
\begin{equation}\label{eq:basis_kappa}
		\psi_{\kappa,n} = N_\kappa \sinh(n\kappa), \quad \phi_{\kappa,n} = N_\kappa \sinh\left( \left(N + 1 -n\right)\kappa \right),
\end{equation}
and the normalisation constant
\begin{equation}
		N^{-2}_\kappa = \sum_{n=1}^{N} \sinh^2(n\kappa) = \dfrac{1}{2}\left(\dfrac{\cosh\left( (N + 1)\kappa \right)\sinh(N\kappa)}{\sinh(\kappa)} - N \right).
\end{equation}
Taking into account Eq.~\eqref{eq:bound_state_k_Ninf}, we obtain that $N_\kappa^{-2}=\mathcal{O}(h^{2N})$ when $N\to\infty$. If we insert Eq.~\eqref{eq:bound_state_k_Ninf} into the dispersion relation~\eqref{eq:tfic_disp}, $\tilde{\varepsilon}(\kappa)=\mathcal{O}(h^{N})$. Thus, the contribution of the boundary state to the correlator~\eqref{eq:obc_ising_corr} is at most of order 
$$ \psi_{\kappa, n}\phi_{\kappa, n} \tanh\left(\dfrac{\beta \lambda(\kappa)}{2}\right)\xrightarrow{N\gg 1} \mathcal{O}(h^N) $$
for $\beta$ finite. Since $0<h<1$, it decays exponentially with system size, and therefore the contribution of the bound state becomes negligible in the thermodynamic limit. The same conclusion follows when working with the alternative basis of \cref{eq:basis2}.

Moreover as $N\rightarrow\infty$ from \cref{eq:quant_cond_2} we have
\begin{equation}
	\dfrac{N}{\pi}k - \dfrac{1}{\pi}\arctan\left(\dfrac{\sin(k)}{\frac{1}{h}-\cos(k)}\right)=n\in\mathbb{Z}
\end{equation}
showing that, apart from the single bound mode, the allowed momenta become uniformly distributed in the interval $(0,\hspace{2pt}\pi)$. Using this and combining \cref{eq:obc_ising_corr} with \cref{,eq:modes_norm} we obtain the two-point correlation functions for a semi-infinite TFIC with a boundary at its
leftmost edge, 
\begin{multline}
		i{\rm Tr}(\rho a^x_l a^y_n)_{\erelbar{21}} = \int_{-\pi}^{\pi}\dfrac{dk}{2\pi}e^{ik(l-n)}\tanh\left(\dfrac{\beta\tilde{\epsilon}(k)}{2}\right)\dfrac{h-e^{-ik}}{\tilde{\epsilon}(k)}   \\
		 - \int_{-\pi}^{\pi}\dfrac{dk}{2\pi}e^{ik(l+n)}\tanh\left(\dfrac{\beta\tilde{\epsilon}(k)}{2}\right)\dfrac{h-e^{-ik}}{\tilde{\epsilon}(k)},\quad l, n=1, 2,\ldots
	\label{eq:obc_ising_corr_sinf}
\end{multline}
which has the characteristic Toeplitz+Hankel structure, with the same symbol as in the infinite TFIC. The open‑boundary effects are encoded in the Hankel term in the second line, while the bound‑state contribution vanishes as $N\rightarrow\infty$.

\subsection{Correlation functions for the baths}
\label[appendix]{app:corr-left-right-bath}
\Cref{eq:obc_ising_corr_sinf} is obtained by taking a chain of $N$ sites with open boundary conditions and subsequently taking $N\rightarrow\infty$. This describes a semi‑infinite system with a boundary at its leftmost edge. Let us rewrite \cref{eq:obc_ising_corr_sinf} as
\begin{equation}
	i{\rm Tr}(\rho a^x_l a^y_n)_{\erelbar{21}} =i{\rm Tr}( a^x_l a^y_n)_{\erelbar{11}} - i {\rm Tr}( a^x_l a^y_{-n})_{\erelbar{11}},  \quad l,n = 1,2,\ldots
	\label{eq:corr-ising-majorana-sinf}
\end{equation}
where 
\begin{equation}
	i{\rm Tr}(\rho a^x_l a^y_n)_{\erelbar{11}} = \int_{-\pi}^{\pi}\dfrac{dk}{2\pi}e^{ik(l-n)}\tanh\left(\dfrac{\beta\tilde{\epsilon}(k)}{2}\right)\dfrac{h-e^{-ik}}{\tilde{\epsilon}(k)}
	\label{eq:corr-ising-majorana-inf}
\end{equation}
is the correlation function of the infinite TFIC in the Gibbs ensemble. That is, the Gibbs ensemble two-point correlation functions in the presence of a boundary at the left-most edge,  ${\rm Tr}(\rho a^x_l a^y_n)_{\erelbar{21}}$, can be decomposed into the Gibbs ensemble correlation for the infinite chain, ${\rm Tr}(\rho a^x_l a^y_n)_{\erelbar{11}}$, minus its \textit{image} ${\rm Tr}(\rho a^x_l a^y_{-n})_{\erelbar{11}}$. Expressing the exponential term $e^{ik(l-n)}$ in \cref{eq:corr-ising-majorana-sinf} in terms of cosine and sine functions, we obtain
\begin{equation}\label{eq:cd-y}
		i{\rm Tr}(\rho a^x_l a^y_n)_{\erelbar{21}} = S_{l-n} + A_{l-n} - S_{l+n} - A_{l+n},
\end{equation}
where
\begin{eqnarray}\label{eq:corr-ising-majorana-sinf-s}
		S_{l-n} = \displaystyle\int_{-\pi}^{\pi}\dfrac{dk}{2\pi}\cos\left( k(l-n) \right)\tanh\left(\dfrac{\beta\tilde{\epsilon}(k)}{2}\right)\dfrac{h-\cos(k)}{\tilde{\epsilon}(k)},\\
		A_{l-n} =  -\displaystyle\int_{-\pi}^{\pi}\dfrac{dk}{2\pi}\sin\left( k(l-n) \right)\tanh\left(\dfrac{\beta\tilde{\epsilon}(k)}{2}\right)\dfrac{\sin(k)}{\tilde{\epsilon}(k)}.
	\label{eq:corr-ising-majorana-sinf-a}
\end{eqnarray}
Note that $S_{l-n} = S_{n-l}$, but $A_{l-n} = A_{n-l}$. From \cref{eq:cd-y}, we obtain
\begin{equation}\label{eq:cd-y2}
	\left[C\right]^{\erelbar{21}}_{ln} = \dfrac{1}{2}\delta_{ln} - \dfrac{1}{2}\left(S_{l-n} -  S_{l+n} - A_{l+n}  \right), \quad \left[F\right]^{\erelbar{21}}_{ln} = \dfrac{1}{2} A_{n-l}, \quad  l,n=1,2,\ldots.
\end{equation}

To compute the correlation function for the left bath, we introduce the mapping
\begin{equation}
	n'=-(N+1)+n \Leftrightarrow n = n' + (N+1),
	\label{eq:map-left-right}
\end{equation}
which takes the lattice sites
$$\{1,2,\ldots,N\}\mapsto\{-N,-(N+1),\ldots,-2,-1\}.$$
In the basis used in \cref{eq:basis1}, this mapping yields
\begin{equation}
	\psi_{k,n'} = N_k \sin \left( \left(N + 1 -|n'|\right)k \right), \quad 
    \phi_{k,n'} = N_k \sin\left(|n'|k\right)
	\label{eq:mapped_basis1}
\end{equation}
for the mode functions.
As in \cref{eq:basis2}, using \cref{eq:helper_thermodyanmic_limit}, we may equivalently adopt the alternative representation
\begin{equation}
	\psi_{k,n'} = \left( N_k/\tilde{\epsilon}(k) \right) \left( h\sin\left(|n'|k\right) - \sin\left(\left(|n'|-1\right)k\right) \right), \hspace{3pt} \phi_{k,n'} = N_k \sin\left(|n'|k\right),
	\label{eq:mapped_basis2}
\end{equation}
which is often more convenient for taking the thermodynamic limit. From \cref{eq:obc_ising_corr,eq:mapped_basis2}, and sending $N\rightarrow\infty$, we arrive at
\begin{equation}
		i{\rm Tr}(\rho a^x_{l'} a^y_{n'})_{\erelbar{12}}  
        = i {\rm Tr}(\rho a^x_{l'} a^y_{n'})_{\erelbar{11}} - i {\rm Tr}(\rho a^x_{-l'} a^y_{n'})_{\erelbar{11}}, \quad l',n'=\ldots,-2,-1. 
	\label{eq:obc_ising_corr_sinf_dbdy}
\end{equation}
Applying Eqs.~\eqref{eq:corr-ising-majorana-sinf-s} and~\eqref{eq:corr-ising-majorana-sinf-a}, we can rewrite it in the form
\begin{equation}
		i{\rm Tr}(\rho a^x_{l'} a^y_{n'})_{\erelbar{12}}= S_{l'-n'} + A_{l'-n'} - S_{-l'-n'} - A_{-l'-n'}. 
\end{equation}
Comparing \cref{eq:corr-ising-majorana-sinf,eq:obc_ising_corr_sinf_dbdy}, we see that for a boundary at the rightmost edge (equivalently, the left boundary pushed to infinity), the reflection acts on the Majorana operator $a^x_{l'}$. This gives for the bath in the left
\begin{equation}
	\left[C\right]^{\erelbar{12}}_{l',n'} = \dfrac{1}{2}\delta_{l',n'} - \dfrac{1}{2}\left(S_{l'-n'} -  S_{-l'-n'} -A_{-l'-n'}  \right), \quad \left[F\right]^{\erelbar{12}}_{l',n'} = \dfrac{1}{2} A_{n'-l'} \quad  l',n'=\ldots,-2,-1.
	\label{eq:cd-y3}
\end{equation}
From \cref{eq:cd-y2,eq:cd-y3} we observe that 
\begin{equation}
	\begin{gathered}
		\left[C\right]^{\erelbar{21}}_{l,n}=\left[C\right]^{\erelbar{12}}_{-l,-n},\\
        \left[F\right]^{\erelbar{21}}_{l,n}=- \left[F\right]^{\erelbar{12}}_{-l,n},\quad  l,n=1,2,\ldots.
	\end{gathered}
\end{equation}
For example, $\left[C\right]^{\erelbar{21}}_{2,5} = \frac{1}{2}\left(S_{3} - S_{7} -A_{7}  \right)=\left[C\right]^{\erelbar{12}}_{-2,-5}$ while $\left[F\right]^{\erelbar{21}}_{2,5}=\frac{1}{2}A_3=-\left[F\right]^{\erelbar{21}}_{-2,-5}$. Hence the anomalous correlation function gets a minus sign on reflection. For the XX chain, the anomalous correlation function $F$ vanishes, and no such subtlety arises.

\section{Time evolution of the two-point correlation matrix in the infinite TFIC}\label{app:ev_TIC}

In this Appendix, we obtain the time evolution
of the two-point correlation matrix under the evolution of the infinite quantum Ising chain Hamiltonian in Eq.~\eqref{eq:Ising_JW_H}. 

To this end, it is convenient to find the modes that diagonalise this Hamiltonian. Performing the Fourier transformation
\begin{equation}
\begin{pmatrix}
		d_k \\ d^{\dagger}_{-k}
	\end{pmatrix} =
	\sum_{n\in \mathbb{Z}} \underbrace{e^{- i k n}}_{\left[ \mathcal{F}^{-1} \right]_{kn}}
	\begin{pmatrix}
		c_n \\ c^{\dagger}_n
	\end{pmatrix}
	\label{eq:XY_Fourier}
\end{equation}
and subsequently the Bogoliubov transformation
\begin{equation}\label{eq:bog_op_infinite_TIC}
\begin{pmatrix}
			\eta_k \\ \eta^{\dagger}_{-k}
		\end{pmatrix} = 
		\underbrace{
			\begin{pmatrix}
				\cos(\Delta_k/2) & i \sin(\Delta_k/2) \\
				i \sin(\Delta_k/2) & \cos(\Delta_k/2)
			\end{pmatrix}
		}_{B_k} 
		\begin{pmatrix}
			d_k \\ d^{\dagger}_{-k}
		\end{pmatrix} 
\end{equation}
with the trigonometric coefficients given by \cref{eq:cdk-sdk-xy} with $\gamma = 1$, the Hamiltonian~\eqref{eq:Ising_JW_H} takes the 
form
\begin{equation}
	H = \sum_k \tilde{\epsilon}(k) \left( \eta^{\dagger}_k \eta_k - \dfrac{1}{2} \right), \quad \tilde{\epsilon}(k) = \sqrt{1 -2h\cos(k)^2 + h^2}.
	\label{eq:JW_XYHamiltonian_diag}
\end{equation}
We emphasise that the operators $\eta_k$ appearing here differ from those used earlier to diagonalise the TFIC with open boundary conditions. Thus, the Heisenberg evolution of the Bogoliubov operators~\eqref{eq:bog_op_infinite_TIC} is simply
\begin{equation}
	\begin{pmatrix}
		\eta_k(t) \\ \eta^{\dagger}_{-k}(t)
	\end{pmatrix} = 
	\underbrace{
		\begin{pmatrix}
			e^{-i \tilde{\epsilon}(k) t} & 0 \\
			0 & e^{i \tilde{\epsilon}(k) t}
		\end{pmatrix}
	}_{\Omega_k(t)}
	\begin{pmatrix}
		\eta_k \\ \eta^{\dagger}_{-k}
	\end{pmatrix},
	\label{eq:eta_Heis}
\end{equation}
where we took into account the symmetry $\tilde{\epsilon}(-k) = \tilde{\epsilon}(k)$.

Let us define the correlation matrix
\begin{equation}
	W_{nm}(t) = \mathrm{Tr} \left[  \rho(t)  
	\begin{pmatrix}
		c_n \\ c^{\dagger}_n
	\end{pmatrix}
	\begin{pmatrix}
		c^{\dagger}_m, c_m
	\end{pmatrix}
	\right]
	\label{eq:2pt_XY_at_t}
\end{equation}
where $\rho(t)$ denotes the time‑evolved density matrix under \cref{eq:Ising_JW_H},
$\rho(t)=e^{-itH}\rho(0)e^{itH}$. Notice that the original correlation matrix in the main text~\eqref{eq:corr-mat} is given by $\Gamma(t)=2W(t)-I$. Combining Eqs.~\eqref{eq:XY_Fourier} and~\eqref{eq:bog_op_infinite_TIC},  we can write it in terms of the Bogoliubov operators,
\begin{equation}
	W_{nm}(t) = \int_{-\pi}^{\pi}\frac{dk\,dq}{(2\pi)^2} \mathcal{F}_{nk} B^{\dagger}_k 
	\mathrm{Tr} \left[  \rho(t)  
	\begin{pmatrix}
		\eta_{k} \\ \eta^{\dagger}_{-k}
	\end{pmatrix}
	\begin{pmatrix}
		\eta^{\dagger}_{q}, \eta_{-q}
	\end{pmatrix}
	\right]
	B_q \mathcal{F}^{*}_{mq}.
\end{equation}
Applying the Heisenberg evolution in Eq.~\eqref{eq:eta_Heis},
\begin{equation}
W_{nm}(t)= \int_{-\pi}^{\pi}\frac{dk\,dq}{(2\pi)^2}\mathcal{F}_{nk} B^{\dagger}_k \Omega_k(t)
	\mathrm{Tr} \left[  \rho(0)
	\begin{pmatrix}
		\eta_{k} \\ \eta^{\dagger}_{-k}
	\end{pmatrix}
	\begin{pmatrix}
		\eta^{\dagger}_{q},  \eta_{-q}
	\end{pmatrix}
	\right]
	\left[ \Omega_q(t) \right]^{\dagger} B_q \mathcal{F}^{*}_{mq},
\end{equation}
and writing the Bogoliubov modes back in terms of the real space creation and annihilation operators, we finally obtain
\begin{equation}
		W_{nm}(t) = \sum_{l,j \in \mathbb{Z}} \mathbb{T}_{nl}(t) W_{lj}(0) \left[ \mathbb{T}^{\dagger}(t) \right]_{jm}\\
	\label{eq:Wnm_time_evol_2},
\end{equation}
where $\mathbb{T}_{nl}(t)$ is the $2\times 2$ matrix
\begin{equation}
    U_{nl}(t)=\int_{-\pi}^{\pi}\frac{dk}{2\pi}\mathcal{F}_{nk} B^{\dagger}_k \Omega_k(t) B_k \left[ \mathcal{F}^{-1} \right]_{kl}.
\end{equation}
Inserting the explicit expressions in Eqs~\eqref{eq:XY_Fourier},~\eqref{eq:bog_op_infinite_TIC}, and~\eqref{eq:eta_Heis} of $\mathcal{F}_{nk}$, $B_k$ and $\Omega_k(t)$, we find that it takes the particular form
\begin{equation}
\mathbb{T}(t) =
		\begin{pmatrix}
			U(t) & V(t) \\
			- \left[ V(t) \right]^{\dagger} & \left[ U(t) \right]^{\dagger}
		\end{pmatrix},
\end{equation}
where the entries $U(t)$ and $V(t)$ are given by
\begin{equation}\label{eq:U_Wnm_Tevol}
	\left[ U(t) \right]_{nl}  = \int_{-\pi}^{\pi} \dfrac{dk}{2 \pi} \cos(k(n-l)) \left( \cos ( \tilde{\epsilon}(k) t )-i \sin ( \tilde{\epsilon}(k) t ) \cos (\Delta_k ) \right)
\end{equation}
and
\begin{equation}
	\left[ V(t) \right]_{nl} = \int_{-\pi}^{\pi} \dfrac{dk}{2 \pi} \sin(k(n-l)) \sin ( \tilde{\epsilon}(k) t ) \sin (\Delta_k ) 
	\label{eq:V_Wnm_Tevol}.
\end{equation}
One can verify that  $$ U(t) U(t)^{\dagger} + V(t) V(t)^{\dagger} = I$$ 
and that $U(t)$, $V(t)$, $U(t)^{\dagger}$, and $V(t)^{\dagger}$ mutually commute. Consequently, the evolution matrix $\mathbb{T}(t)$ satisfies 
$$\mathbb{T}(t) \left[ \mathbb{T}(t) \right]^{\dagger} = \mathbb{I}$$ 
as expected. From Eqs.~\eqref{eq:U_Wnm_Tevol} and \eqref{eq:V_Wnm_Tevol}, we further observe that $U(t)$ is symmetric, $U(t)^T = U(t)$, and $V(t)$ antisymmetric, $V(t)^T=-V(t)$.

\end{document}